\documentclass[twocolumn]{aastex63}

\received{December 22, 2019}
\revised{February 27, 2020}
\accepted{March 9, 2020}

\newcommand{\teff}{$T_\mathrm{eff}$}
\newcommand{\logg}{$\log g$}
\newcommand{\feh}{\textrm{[Fe/H]}}
\newcommand{\micro}{$\xi_\mathrm{micro}$}

\newcommand{\kms}{km\,s$^{-1}$}
\newcommand{\mic}{$\mu \mathrm m$}
\newcommand{\msun}{$M_\odot$}

\shorttitle{Fluorine in the Galaxy}
\shortauthors{Ryde et al.}
\graphicspath{{./}{figures/}}

\begin{document}

\title{FLUORINE IN THE SOLAR NEIGHBORHOOD: THE NEED FOR SEVERAL COSMIC SOURCES} 

\correspondingauthor{Nils Ryde}
\email{ryde@astro.lu.se}

\author[0000-0001-6294-3790]{Nils Ryde}
\affiliation{Lund Observatory, Department of Astronomy and Theoretical Physics, 
Lund University, Box 43, SE-221 00 Lund, Sweden}

\author[0000-0002-4912-8609]{Henrik J\"onsson}
\affiliation{Materials Science and Applied Mathematics, Malm\"o University, SE-205 06 Malm\"o, Sweden}
\affiliation{Lund Observatory, Department of Astronomy and Theoretical Physics, Lund University, Box 43, SE-221 00 Lund, Sweden}

\author[0000-0001-7875-6391]{Gregory Mace}
\affiliation{Department of Astronomy and McDonald Observatory, The University of Texas, Austin, TX 78712}

\author[0000-0001-6476-0576]{Katia Cunha}
\affiliation{Observat\'orio Nacional San Crist\'ovao, Rio de Janeiro, Brazil}
\affiliation{University of Arizona, Tucson, AZ 85719, USA}

\author[0000-0001-9715-5727]{Emanuele Spitoni}
\affiliation{Stellar Astrophysics Centre, Department of Physics and Astronomy, Aarhus University, Ny Munkegade 120, DK-8000 Aarhus C,
Denmark}


\author[0000-0002-2516-1949]{Melike Af\c{s}ar}
\affiliation{Department of Astronomy and Space Sciences, Ege University, 35100 Bornova, \.Izmir, Turkey}

\author[0000-0003-3577-3540]{Daniel Jaffe}
\affiliation{Department of Astronomy and McDonald Observatory, The University of Texas, Austin, TX 78712}


\author[0000-0001-6079-8630]{Rebecca Forsberg}
\affil{Lund Observatory, Department of Astronomy and Theoretical Physics, 
Lund University, Box 43, SE-221 00 Lund, Sweden}

\author[0000-0001-6909-3856]{Kyle F. Kaplan}
\affiliation{SOFIA Science Center, Universities Space Research Association, NASA Ames Research Center, MS 232-12, Moffett Field, CA 94035, USA}

\author[0000-0002-2039-7143]{Benjamin T. Kidder}
\affiliation{Department of Astronomy and McDonald Observatory, The University of Texas, Austin, TX 78712}

\author[0000-0003-0894-7824]{Jae-Joon Lee}
\affiliation{Korea Astronomy and Space Science Institute 776, Daedeok-daero, Yuseong-gu, Daejeon, 34055, Republic of Korea}

\author[0000-0002-0418-5335]{Heeyoung Oh}
\affiliation{Korea Astronomy and Space Science Institute 776, Daedeok-daero, Yuseong-gu, Daejeon, 34055, Republic of Korea}

\affil{Department of Astronomy, University of Texas at Austin, Austin, TX 78712, USA}

\author[0000-0002-0134-2024]{Verne V. Smith}
\affiliation{National Optical Astronomy Observatories, Tucson, AZ 85719, USA}

\author[0000-0002-3456-5929]{Christopher Sneden}
\affiliation{Department of Astronomy and McDonald Observatory, The University of Texas, Austin, TX 78712}

\author{Kimberly R. Sokal}
\affiliation{Department of Astronomy and McDonald Observatory, The University of Texas, Austin, TX 78712}

\author{Emily Strickland}
\affiliation{Department of Astronomy and McDonald Observatory, The University of Texas, Austin, TX 78712}

\author[0000-0002-5633-4400]{Brian Thorsbro}
\affil{Lund Observatory, Department of Astronomy and Theoretical Physics,
Lund University, Box 43, SE-221 00 Lund, Sweden}



\begin{abstract}

The cosmic origin of fluorine is still not well constrained. Several nucleosynthetic channels at different phases of stellar evolution have been suggested, but these must be constrained by observations. For this, the fluorine abundance trend with metallicity spanning a wide range is required. Our aim is to determine stellar abundances of fluorine for $-1.1<\feh<+0.4$. We determine the abundances from HF lines in infrared K-band spectra ($\sim 2.3\,$\micron) of cool giants, observed with the IGRINS and Phoenix high-resolution  spectrographs. 
We derive accurate stellar parameters for all our observed K giants, which is important since the HF lines are very temperature sensitive. We find that [F/Fe] is flat as a function of metallicity at [F/Fe]$\sim 0$, but increases as the metallicity increases. The fluorine slope shows a clear secondary behavior in this metallicity range. We also find that the [F/Ce] ratio is relatively flat for $-0.6<\feh<0$, and that for two metal-poor  ($\feh<-0.8$),  s-process element enhanced giants,  we do not detect an elevated fluorine abundance. We interpret all these observational constraints to indicate that several major processes are at play for the cosmic budget of fluorine over time; from those in massive stars at low metallicities, through the asymptotic giant branch-star contribution at $-0.6<\feh<0$, to processes with increasing yields with metallicity at super-solar metallicities. The origins of the latter, and whether or not Wolf-Rayet stars and/or novae could contribute at super-solar metallicities, is currently not known. To quantify 
these observational results, theoretical modelling is required. More observations in the metal-poor region are required to clarify the processes there.





\end{abstract}



\vskip 10mm

\section{Introduction} \label{sec:intro}


The cosmic origin of fluorine, i.e. the sites and processes that are responsible for the build-up of the element in the Universe, but also its galactic chemical evolution, are still very uncertain and turn out to be very intriguing. The solar abundance of fluorine is less than a percent 
of that of the neighboring elements in the periodic table, which is a reflection of its unique formation channels.  
Fluorine  reacts readily  with hydrogen and helium in stellar interiors, via the reactions $^{19}$F($p,\alpha$)$^{16}$O and $^{19}$F($\alpha,p$)$^{22}$Ne, which destroy the $^{19}$F nuclei. There are, however, several possible nucleosynthetic reaction chains acting in different evolutionary processes of stars that can synthesize the fragile fluorine nuclei so that they survive and contribute to the build-up of the cosmic reservoir of fluorine. It is debated which of the possible processes is dominant in the Universe and the relative importance of them at different epochs. Measuring the fluorine abundances as a function of time or metallicity provides important constraints to the different formation channels and  different theories of the formation of fluorine.

Several of the theoretically suggested processes \citep[see, e.g., the discussion in][]{spitoni:18} could actually, within their uncertainties and reasonable ranges of input parameters, by themselves produce all of the measured cosmic fluorine abundance. However, the different processes act on different timescales, which means that the evolution of the build-up of fluorine will be very different. Observational constraints on these evolutionary trends will test the importance of the processes. The observed trends might also reflect several processes and might be different for different stellar populations (for example, the thin-disk, thick-disk, and bulge populations). 
At low metallicities, the $\nu$ process \citep{woosley:88} and the contribution from rapidly-rotating, massive stars \citep{prantzos:18} can be tested. At solar metallicities the contribution from the thermally-pulsating asymptotic giant branch (TP-AGB) stars \citep{jorissen:92} will be the largest, and at higher metallicities, and especially at super-solar metallicities, the contributions from novae \citep{spitoni:18},  Wolf-Rayet stars \citep{meynet:00}, or from some other metallicity-dependent process might become increasingly strong. Whether or not Wolf-Rayet stars actually would contribute to the cosmic budget of fluorine is, however, highly uncertain (\citet{palacios:05} and G. Meynet, private communication). 

However, determining the fluorine abundance is challenging. There is only one stable isotope, $^{19}$F, and no useful atomic lines are readily available for an abundance determination in cool, stellar atmospheres. The highly ionized lines in the far-UV \citep{werner:05} and the highly excited F{\sc i} lines at 6800-7800\,\AA\ are only observed in hot stars. The latter were used in Extreme Helium Stars and R Coronae Borealis stars \citep{pandey:06,pandey:08}, with temperatures of \teff$>6500$\,K.  
The only readily useful diagnostics are lines from  the HF molecule in the K and N bands ($2.1-2.4\,$\micron\ and $8-13\,$\micron, respectively), lines which are  observable only in cool giants (\teff$<4500$\,K). In the K band, telluric lines can render the abundance determinations uncertain \citep{delaverny:13}. The molecular lines are also sensitive to the effective temperatures of the stars, which therefore have to be determined with high accuracy. The diagnostically interesting metal-poor region (investigating the role of and yields from rapidly rotating massive stars and/or the $\nu$ process) is very difficult to address observationally. The HF line lists, both for vibration-rotation lines in the K band, as well as the pure rotational lines in the N band, including the needed partition functions, are now well determined, see the discussions in \citet{jonsson:14,jonsson:14b}. 

The field of observationally investigating the chemical evolution of fluorine has grown in recent years due to the advent of sensitive, high resolution spectrometers recording light in the infrared; for example, \citet{recio:12,jonsson:14b} used CRIRES at the Very Large Telescope  and  \citet{pila:15,jonsson:17,rafael:19,rafael:19:M4} used the Phoenix spectrometer to measure fluorine abundances in K-band spectra. \citet{rafael:19} also used iSHELL at the NASA Infrared Telescope Facility (IRTF). Furthermore, \citet{jonsson:14} used TEXES spectra to measure abundances from the rotational lines at $12$\,\mic\  as well as archival, K-band spectra observed with the  Fourier Transform Spectrometer at Kitt Peak National Observatory. Theoretical work during the most recent years include those of \citet{prantzos:18,spitoni:18,olive:19}, apart from work done on the nucleosynthetic reaction rates \citep{sieverding:18,sieverding:19,langanke:19}.  A more detailed discussion of these recent investigations and their interpretation will be given in Section \ref{discussion}. The investigation of the cosmic budget of fluorine is very active and  will still require more observational and theoretical work in the future.

Here, we analyse the fluorine abundances in 61 stars, with carefully and homogeneously determined stellar parameters. The latter is important to minimise the systematic uncertainties inherent of the used HF line.  In this way we will be able to provide the largest set of homogeneously determined fluorine abundances for a range of metallicities.  We analyse new K-band spectra observed with the Immersion GRating INfrared spectrograph \citep[IGRINS;][]{igrins,igrins:14} and re\-analyse K-band spectra observed with the Phoenix spectrograph, using more accurate stellar parameters.  The spectra from both instruments have very similar spectral resolving powers and signal-to-noise ratios.  The purely rotational HF lines at $12\,\micron$, presented in \cite{jonsson:14}, are stronger than the vibration-rotation lines at $2.3\,\micron$ (K band), and should be explored further in the future for an abundance investigation of cool giants in the metal-poor region.
Observations at $12\,\micron$ require, however, brighter stars due to insufficiently sensitive spectrographs and less light from stars in the N band \citep{jonsson:14}.

\begin{figure}
\centering
\epsscale{1.00}
\includegraphics[trim={0.2cm 0cm 0cm -1cm},clip,angle=0,width=1.00\hsize]{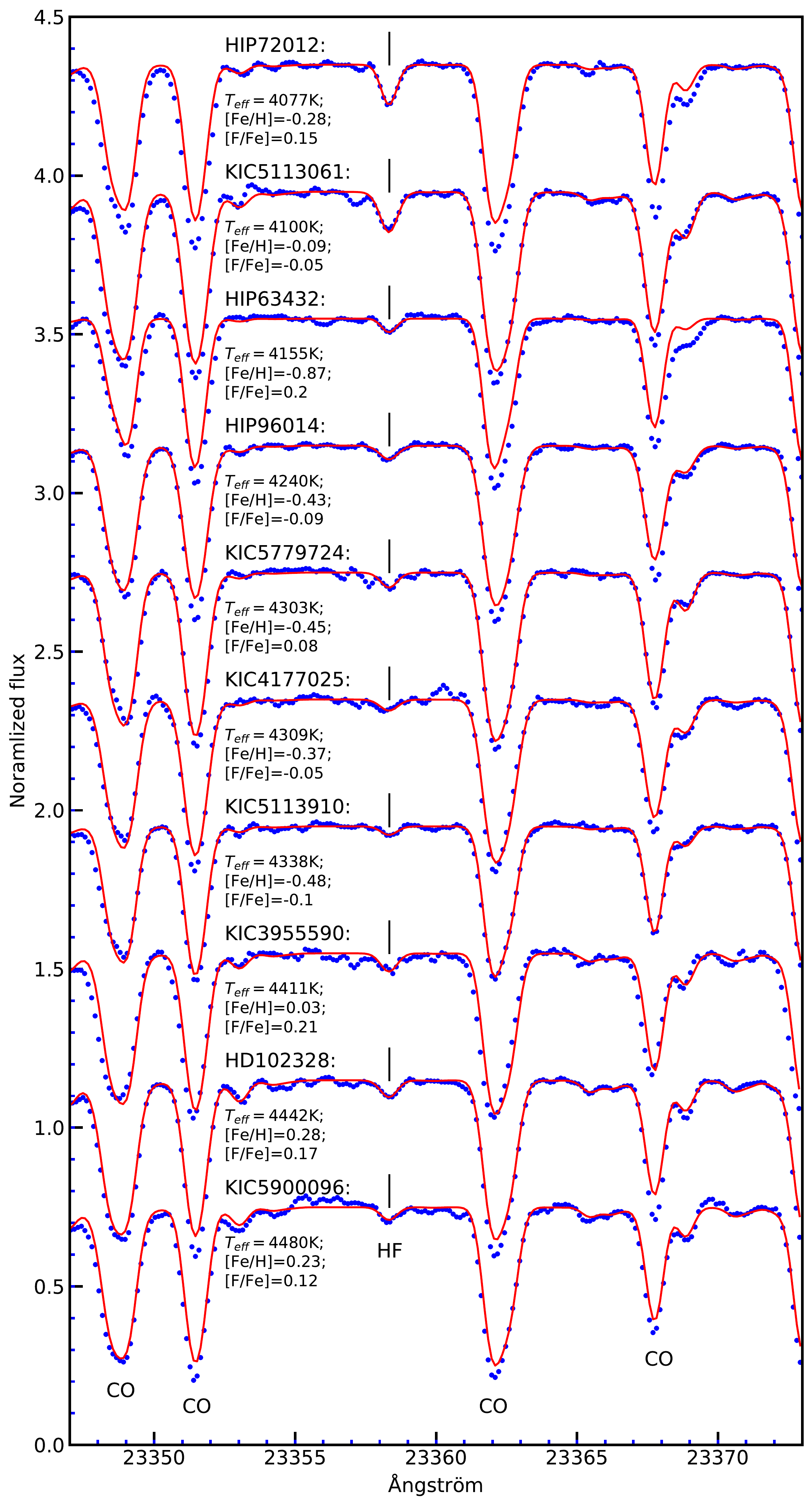} 
\caption{Observed IGRINS spectra of the 10 stars yielding a fluorine abundance. The spectra near the HF line at $\lambda_\mathrm{air}= 23 358.33$\,\AA, is shown with blue dotted lines. Synthetic spectra are shown by the red line. The HF line is marked with vertical lines. Only the HF line is fitted. The stars are ordered with increasing \teff\ from the top. The other spectral lines in the figure are all CO vibration-rotational lines. \label{fig:HF}}
\end{figure}

\section{Observations and Data Reduction} \label{sec:obs}

\begin{deluxetable*}{llCCllL}
\tablecaption{IGRINS observing log in the same order as in Table \ref{tab:obs} \label{tab:logg}}
\tablewidth{0pt}
\tablehead{
\colhead{Star} & \colhead{2MASS name} & \colhead{$H$}  & \colhead{$K$}   & \colhead{Date} & \colhead{Telescope}    & \colhead{Exposure time}\\
\nocolhead{ } & \nocolhead{} & \colhead{(mag)}  & \colhead{(mag)} & \nocolhead{ } & \nocolhead{ }    & \colhead{[s]} 
}
\decimalcolnumbers
\decimals
\startdata
\multicolumn{3}{l}{Stars with a detected HF line:}\\
HIP72012   & J14434444+4027333 & 2.6  & 2.4   & 2016 June 16  & HJST  & 30\times 4\, \mathrm{(ABBA)} \\ 
KIC5113061 & J19413439+4017482 & 8.2  & 8.0   & 2016 Nov. 22  & DCT            & 90\times 8\, \mathrm{(ABBAABBA)} \\ 
HIP63432   & J12595500+6635502 & 2.4  & 2.1   & 2016 May 29  & HJST  & 30\times 6\,  \mathrm{(ABBAAB)} \\ 
HIP96014   & J19311935+5018240 & 2.9  & 2.5   & 2016 June 15  & HJST  & 30\times 4\, \mathrm{(ABBA)}  \\ 
KIC5779724 & J19123427+4105257 & 8.0  & 7.8   & 2016 Dec. 09  & DCT            & 60 \times 10\,\mathrm{(ABBAABBAAB)}   \\ 
KIC4177025 & J19434309+3917436 & 7.6  & 7.5   & 2016 Nov. 22  & DCT            & 60\times 6\,  \mathrm{(ABBAAB)}  \\ 
KIC5113910 & J19421943+4016074 & 8.2  & 8.0   & 2016 Nov. 22  & DCT            & 90\times 8\, \mathrm{(ABBAABBA)}  \\ 
KIC3955590 & J19272677+3900456 & 7.8  & 7.7   & 2016 Nov. 23  & DCT            & 60\times 8\, \mathrm{(ABBAABBA)}  \\ 
HD102328   & J11465561+5537416 & 2.9  & 2.6   & 2016 Feb. 02  & HJST  & 1.6\times 20\, \mathrm{(A\,\&\,B)}  \\
KIC5900096 & J19515137+4106378 & 6.0  & 5.8   & 2016 Nov. 22  & DCT  & 30\times 4\, \mathrm{(ABBA)} \\ 
 & \\
\multicolumn{3}{l}{Stars yielding an upper limit of the HF abundance:}\\
HIP50583          & J10195836+1950290 & -0.8 & -0.8  &  2016 June 20 & HJST  & 1.6\times 18\, \mathrm{(A)}   \\ 
KIC11045542       & J19530590+4833180 &  8.4 &  8.2  & 2016 Dec. 11 & DCT & 250\times  8\, \mathrm{(ABBAABBA)}   \\ 
$\alpha$ Boo      & J14153968+1910558 & -2.8 & -2.9  &  2015 April 11 & HJST  & 30\times 2\, \mathrm{(AB)}  \\ 
2M14231899        & J14231899+0540079 &  8.0 &  7.8 &  2016 June 19 & HJST  & 150\times 8\, \mathrm{(ABBA)}   \\ 
2M17215666        & J17215666+4301408 &  7.6 &  7.5  & 2016 July 25 & HJST  & 180\times 6\, \mathrm{(ABBAAB)}   \\ 
KIC4659706        & J19324055+3946338 &  7.6 &  7.4  & 2016 Nov. 19 & DCT & 60\times 8\, \mathrm{(ABBAABBA)}   \\ 
HIP90344          & J18255915+6533486 &  2.2 &  2.1 &  2016 June 15 & HJST  & 30\times 4\, \mathrm{(ABBA)}   \\ 
KIC3936921        & J19023934+3905592 &  8.3 &  8.1  & 2016 Nov. 23& DCT & 120\times 8\, \mathrm{(ABBAABBA)}   \\ 
KIC11342694       & J19110062+4906529 &  7.6 &  7.4 & 2016 Nov. 17 & DCT & 60\times 8\, \mathrm{(ABBAABBA)}   \\ 
KIC3748585        & J19272877+3848096 &  6.4 &  6.3  & 2016 Nov. 17& DCT & 30\times 8\, \mathrm{(ABBAABBA)}   \\ 
\enddata
\tablecomments{DCT: the Discovery Channel Telescope, a 4.3 m telescope at Lowell Observatory, Arizona. }
\tablecomments{HJST: the  Harlan J Smith Telescope, a 2.7 m telescope at McDonald Observatory, Texas. }
\end{deluxetable*}

Our goal is to fill in and expand the fluorine trends presented in \citet{jonsson:17} for $-0.5<\feh<0.4$, especially expanding the metallicity range both downward and upward. We have selected such stars from a careful optical analysis of about 500 giants to be presented in J\"onsson et al. (in prep.).
These stars are all warmer than 4000 K (K giants), and we therefore avoid the AGB stars that can produce fluorine themselves and pollute their atmospheres \citep[see, e.g.,][]{jorissen:92}. Hence, all our stars are useful probes for the galactic chemical evolution of fluorine. 

Since the fluorine abundances are determined from vibration-rotational lines of the HF molecule, 
stars that yield suitable HF-line strengths have to be chosen. 
For a given metallicity, these molecular lines become stronger 
the cooler the star is and the lower the surface gravity of the star is;
the lower the temperature is, the larger is the molecular
density, and the lower the surface gravity is the stronger is
the relative strength of the lines  compared to the
continuum. The continuous opacity in this wavelength region, which is due to the $H^-_\mathrm{ff}$ process, decreases with electron pressure in the line-forming regions, which in turn decreases with lower surface gravities. Since the line strengths are proportional to the ratio of line to continuous opacities, the lines become stronger for a star with a lower surface gravity, for a given fluorine abundance \citep[see][]{jonsson:14b}.  Thus, in order for the lines to be measurable in stars of low metallicities, cool giants should to be chosen. In the metal-rich wavelength region, the lines can even become saturated for very cool giants. 

Spectra of giants from two sets of observations have been analysed here:
the first set consists of 25 giants, which were observed with the IGRINS spectrograph \citep{igrins}.
Of these giants, 
10 had spectra with a detectable HF feature and another 10 yielded a useful upper limit. The second set consists of 41 giants with spectra from \citet{jonsson:17} displaying a detectable HF line and are reanalysed here.  These were observed with the Phoenix spectrograph \citep{phoenix,phoenix:2003} mounted on the 4m Mayall telescope at Kitt Peak National Observatory (KPNO) at a spectral resolving power of $R=50000$ and typical signal-to-noise ratios of 100 \cite[for more details, see, ][]{jonsson:17}.

We have also used the derived abundances from six giants from the work by \citet{jonsson:14}. They observed these stars with the TEXES spectrograph \citep{texes} at NASA's Infrared Telescope Facility (IRTF), recording the rotational HF lines at $12\,\micron$  with a spectral resolving power of $R\sim65\,000$ and a signal-to-noise ratio also of typically 100. They used a similar method of analysis as presented here.

The IGRINS spectra were recorded during 2016, from February to December, apart from $\alpha$ Boo which was observed earlier on 2015 April 11, see Table \ref{tab:logg}. The spectra were all observed on the 4.3-meter Discovery Channel Telescope (DCT) at Lowell Observatory \citep{mace:18}, or on the 2.7 meter Harlan J. Smith Telescope at McDonald Observatory \citep{mace:16}. IGRINS provides a spectral resolving power of $R=\lambda/\Delta\lambda\sim 45\,000$ spanning the full H and K bands ($1.45-2.5\,\micron$), recorded in one exposure, even though for this paper we use only small parts of the spectra. 

The stars were observed in an ABBA nod sequence along the slit. Exposure times for these bright objects range from $30$ to $2000$ seconds, see Table \ref{tab:logg}. These exposure times were set by the requirement to retrieve spectra of a signal-to-noise ratio of at least 100, which was achieved. Telluric standard stars (typically rapidly rotating, late B to early A dwarfs) were also observed in conjunction with the science targets at similar air masses. All the spectra were reduced using the IGRINS pipeline \citep{igrins_pipeline:17}, which extracts wavelength calibrated spectra after flat-field correction and A-B frame subtractions. 

The science spectra were then divided by the telluric spectra, in order to divide out the telluric lines. This works very well, since the telluric stars are observed close in time and at a similar airmass compared to the observations of the science targets. Every order of the divided spectra is continuum normalized with the IRAF task {\tt continuum} \citep{IRAF}. These were then combined with the task {\tt scombine} allowing an addition of overlapping regions of subsequent orders, but also cutting away edge regions with no traceable continuum and spurious edge effects. 
This resulted in one normalized stitched spectrum for the 
K band with a wavelength coverage of $19700-24800$\,\AA. The regions with heavy telluric contamination at the edges of these limits are, however, not always useful. In the cases where the final spectra still have some modulation in their continuum levels, these are taken care of by defining specific local continua around the spectral line being studied. The HF line that is finally used lies at $\lambda_\mathrm{air}= 23 358.33$\,\AA. The 10 spectra with a detected HF line
are shown in Figure \ref{fig:HF}, where the spectra are ordered by increasing \teff. The effective temperatures and  metallicities of the stars are indicated in the figure.


\section{Analysis} \label{sec:anal}

\begin{figure}[ht!]
\centering
\epsscale{1.00}
\includegraphics[trim={0.2cm 0cm 0cm 0cm},clip,angle=0,width=1.00\hsize]{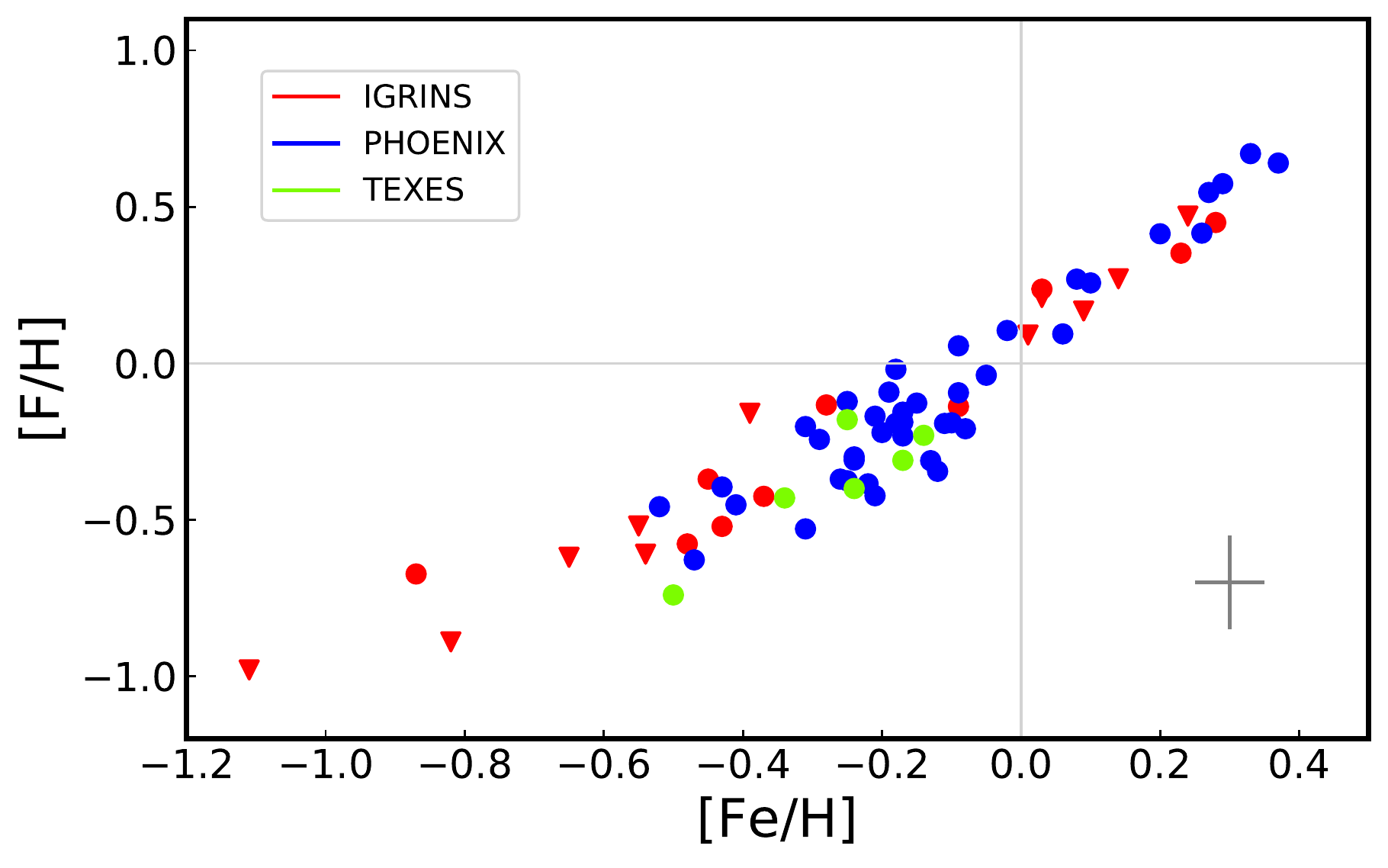}
\caption{[F/H] as a function of metallicitiy, [Fe/H] is shown for the stars observed with IGRINS (red), Phoenix (blue), and TEXES (green).  A(F)$_\odot = 4.43$ \citep{meteoritic:03}. \label{fig:fh_vs_feh}}
\end{figure}

\begin{figure}[ht!]
\centering
\epsscale{1.00}
\includegraphics[trim={0.2cm 0cm 0cm -1cm},clip,angle=0,width=1.00\hsize]{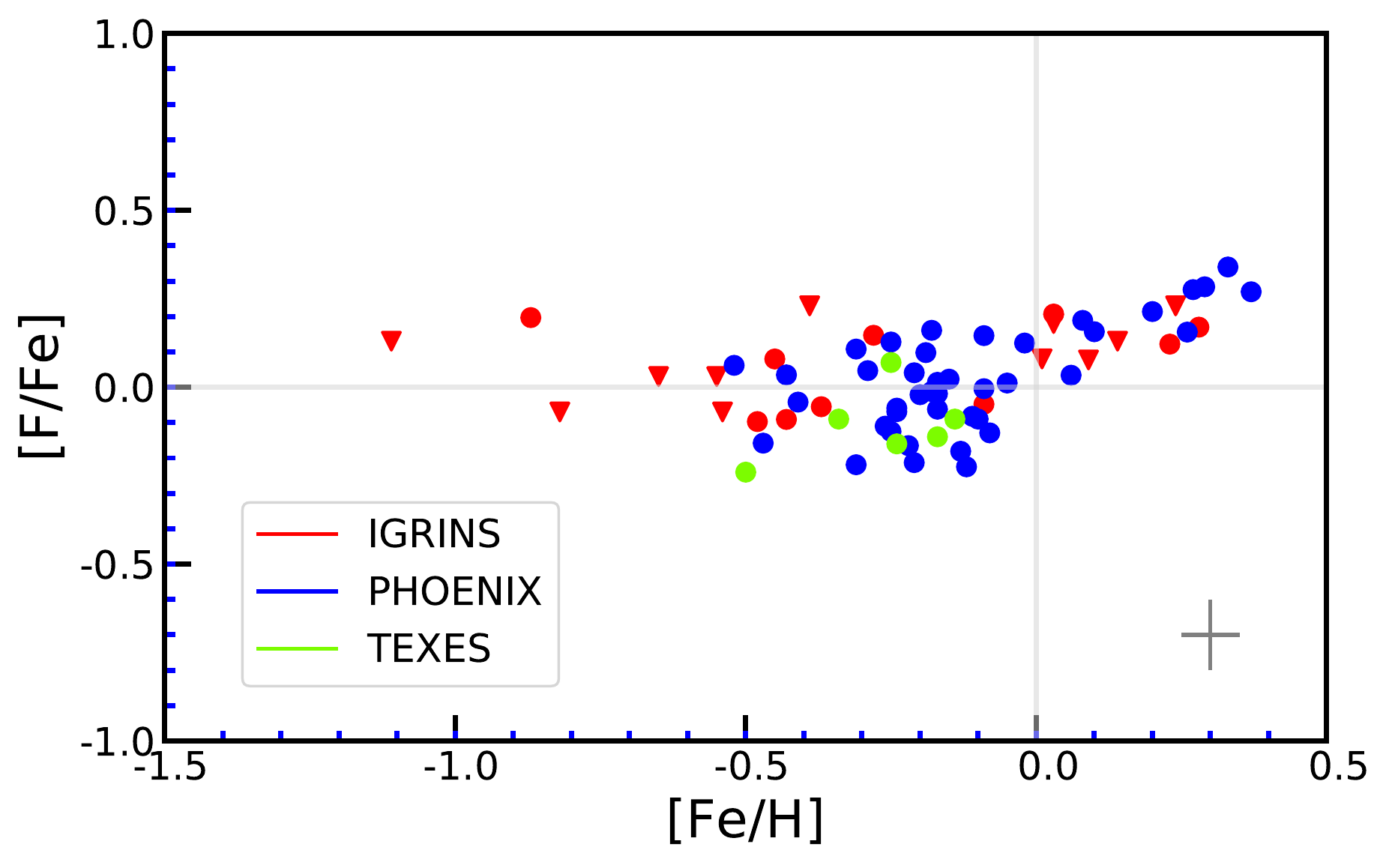} 
\caption{[F/Fe] ratio as a function of metallicity is shown for the stars observed with IGRINS (red), Phoenix (blue), and TEXES (green). A(F)$_\odot = 4.43$ \citep{meteoritic:03}. \label{fig:ffe_vs_fe}}
\end{figure}

From the IGRINS spectra of these 10 new stars and the 41 Phoenix spectra, we have thus derived the fluorine abundance from the HF($v=1-0$) R9 line at $\lambda_\mathrm{air}= 23 358.33$\,\AA. We have analyzed these spectra with tailored synthetic spectra, calculating the radiative transfer through spherical model atmospheres, defined by their stellar parameters. These are the effective temperature, \teff, surface gravity, \logg, 
metallicity, $\feh$, and the microturbulence, \micro.

In order to derive as accurate abundances as possible, these fundamental input parameters must be determined accurately and in a homogeneous way. The spectroscopic method developed for K giants from high-resolution optical spectra by \citet{jonsson:17I} can do that. In this method the stellar parameters are determined simultaneously by fitting unsaturated and unblended Fe I, Fe II and Ca I lines as well as log g sensitive Ca I wings. The derived parameters are benchmarked against independently determined effective temperatures, \teff, from angular diameter measurements and surface gravities, \logg, from asteroseimological measurements. As a development of this method, J\"onsson et al. (in prep.) have been utilizing an up to three times broader wavelength range of the high-resolution optical spectra from \citet{jonsson:17I}, \citet{lomaeva:19}, and \citet{forsberg:19},  all observed with the FIES spectrograph \citep{fies} on the Nordic Optical Telescope (NOT). In total, stellar parameters and several abundances for more than 500 K-giants have been derived. We have been using a subset of these here.

The code Spectroscopy Made Easy (SME) \citep{sme,sme_code} is used to determine these stellar parameters. SME interpolates in a grid of one-dimensional (1D) MARCS atmosphere models \citep{marcs:08}. These are hydrostatic model atmospheres in spherical geometry, computed assuming LTE, chemical equilibrium, homogeneity, and conservation of the total flux (radiative plus convective, the convective flux being computed using the mixing-length recipe). The uncertainties achieved are $\pm 50$\,K for \teff, $\pm0.15$\,dex for \logg, $\pm0.05$ dex for [Fe/H], and $\pm0.1$\,\kms for \micro.  
The final stellar parameters are given in Tables \ref{tab:obs}  and \ref{tab:phoenix}. 

From the optical spectra we have also determined the oxygen and cerium abundances, listed in Tables \ref{tab:obs} and \ref{tab:phoenix}. The oxygen abundances were determined from the $6300.308 $\,\AA\ [O{\sc i}]-line, and the cerium abundance from five Ce II lines between $5250$ and $6050$\,\AA\ (more details in J\"onsson et al. (in prep)). These abundances are used in the trend plots in Figures \ref{fig:AF_vs_AO}-\ref{fig:cef_vs_feh}.


With the derived stellar parameters, we can then synthesize spectra and determine abundances from the infrared spectra.
We have chosen also here to use SME in order to be consistent with the determination of the stellar parameters, but also since it has a flexible chi-square minimization tool for finding the solution that fits an observed spectrum the best, in a pre-specified spectral window. Then, we determine the fluorine abundance by fitting the HF line. To improve the rough normalisation that was initially done on the spectra, a straight line was fitted to continuum regions on both sides of the HF line. Also, the width of the line, which we call $\xi_\mathrm{macro}$ and includes both the stellar macroturbulence and the spectrograph's instrumental profile, is carefully determined since the entire line profile is fitted. For the IGRINS spectra, the $\xi_\mathrm{macro}$ is determined from a few blend-free Si lines with suitable strength for the width determination, and for the narrower Phoenix spectra, the $\xi_\mathrm{macro}$ is determined from the HF features themselves and checking against the neighboring CO lines. A $\xi^{\mathrm{FWHM}}_\mathrm{macro}\sim4.5\,$\kms\ is found for all stars.

The fluorine abundance is thus determined for the 41 giants observed with the Phoenix spectrograph and the 10 giants observed with the IGRINS spectrograph, that show clear HF lines. We have synthesized eight of the most promising HF lines, namely the HF($\nu=1-0$) R3, 4, 7, 9, 12, 13, 14, and 15 of the R branch. However, it is only the R9 line that can be used for an abundance determination; all the other lines have various problems, such as line blending or being too weak. Indeed, it is mostly the HF($\nu= 1 - 0$) R9 line that has been used in the literature.   

The molecular line data, i.e. wavelengths, excitation energies, and transition probabilities ($\log gf$), are calculated and given in \citet{jonsson:14b}. These authors also stress the importance of using a partition function consistent with these excitation energies in order to get the correct abundances; otherwise an abundances off-set of $\sim0,3$ dex is found. The correct partition function is given in \citet{jonsson:14}. For such a light molecule as HF, the difference in energies is large, depending on whether the zero point of the excitation energies is set by the dissociation energy of the energy potential, $D_e$, or the true energy required for dissociation, $D_0$. \citet{jonsson:14b} use the latter definition. It should also be noted that there is still an uncertainty in the dissociation energy of the HF molecule, which could give an additional systematic uncertainty in the derived F abundances of $\sim0.04$ dex \citep[see discsussion in][]{rafael:19}.

In Figure 3 of \citet{jonsson:14b} all vibration-rotational lines in both the R and P branches are given. The P branch lines are in general stronger than the lines in the R branch, with the P10 line being the strongest. However, the P branch lies mainly between the K and L bands, where it is obscured by the Earth's atmosphere. The HF($\nu=1 - 0$) P21 (and higher) lines appear in the L band ($\sim 3.5-4.1\,$\micron). 

For the 10 giants for which a Gaussian line profile could not be distinguished from the noise in the spectra, an upper limit of the fluorine abundance is determined instead. This was done such that a synthetic spectrum with a clearly too strong HF line than the observed noise in the region of the HF line, is calculated, providing an upper limit to the fluorine abundance. Five other giants that were hotter, provided upper limits that were uninterestingly high and are omitted from the analysis for clarity.

\section{Results} \label{sec:res}

\subsection{Fluorine Abundances}

The final fluorine abundances and upper limits are given in Tables \ref{tab:obs} and \ref{tab:phoenix}, where we provide the number density abundances, $A(F)=\log N_F/N_H+12$ and the [F/Fe]{\footnote{ The notation [A/B] = $\log (N_\mathrm{A}/N_\mathrm{B})_* - \log (N_\mathrm{A}/N_\mathrm{B})_\odot$, where $N_\mathrm{A}$ and $N_\mathrm{B}$ are the number abundances of elements A and B respectively.}} abundance ratios. In the tables we also provide the oxygen and cerium abundance results from the optical spectra.

\begin{figure}[ht!]
\centering
\epsscale{1.00}
\includegraphics[trim={0.2cm 0cm 0cm 0cm},clip,angle=0,width=1.00\hsize]{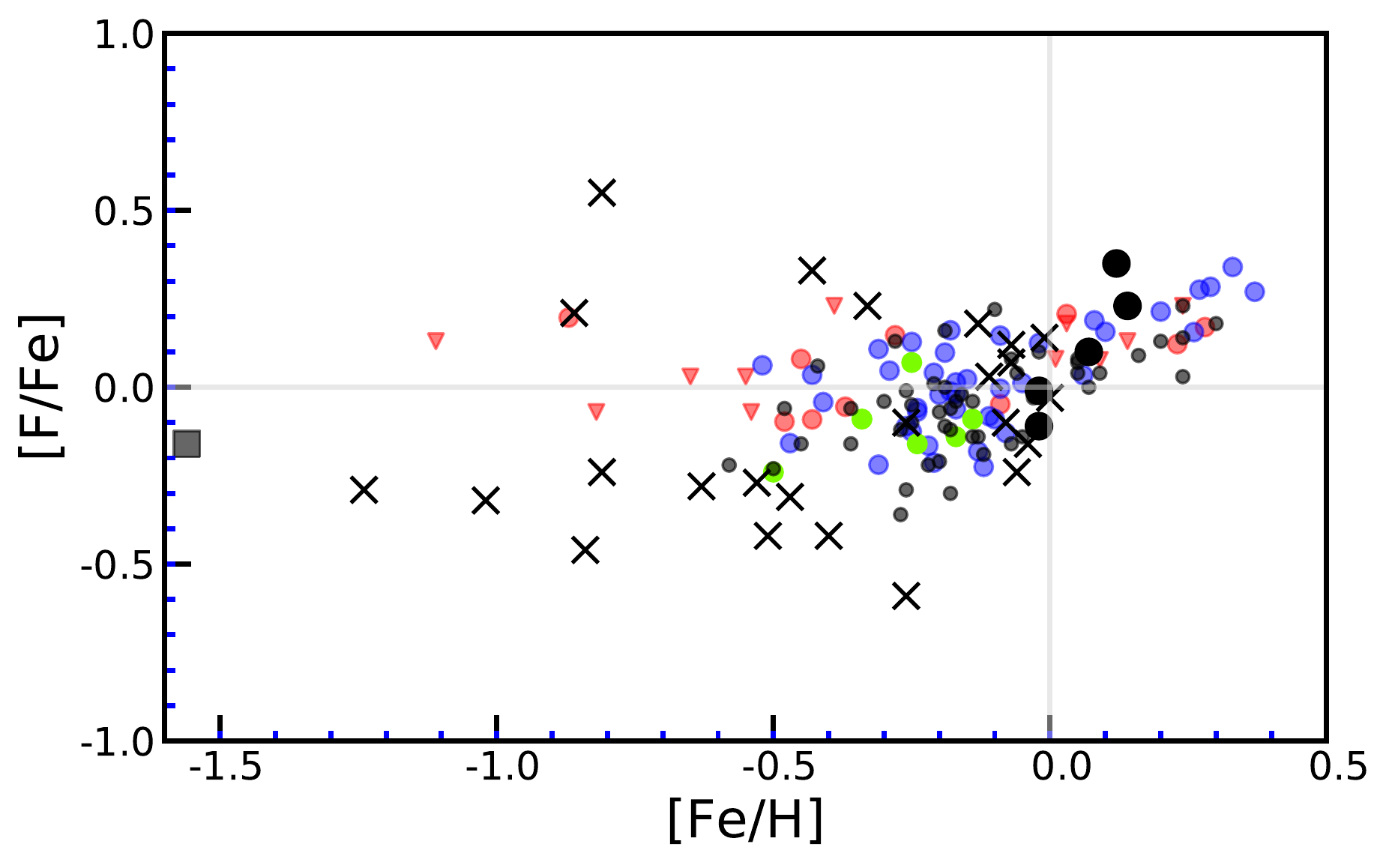}
\caption{[F/Fe] versus [Fe/H] compared to determinations from the literature. Large black dots are from \citet{nault:13}, small grey dots from \citet{jonsson:17}, the  square from \citet{li:13}, and crosses from \citet{rafael:19}, with the two metal-poor stars with high [F/Fe] probably being members of the Monoceros overdensity \citep{rafael:19}. \label{fig:others}}
\end{figure}

In Figure \ref{fig:fh_vs_feh} we present [F/H] versus [Fe/H] as determined from the IGRINS data (in red), the Phoenix data (in blue), and from the \citet{jonsson:14} TEXES determinations (in green). Upper limits are marked with triangles. In Figure \ref{fig:ffe_vs_fe} the [F/Fe] trend is shown instead. The stellar abundances are normalized to the solar value, which is very uncertain; there is no detectable HF in the solar photosphere, so the solar abundance value is determined either from meteoritic measurements \citep[$A_\odot(F)=4.43\pm0.06$;][]{meteoritic:03} or from uncertain measurements in sunspot spectra \citep[$A_\odot(F)=4.56\pm 0.30$;][]{hall:69,solar:sme}\footnote{See also the discussion in \citet{nault:13} about this value and a reevaluation of it due to the problems with the excitation potential of HF, now solved \citep{jonsson:14b}}. More recently \citet{maiorca:14} analysed a spectrum of a medium-strong sunspot umbra ($\sim4250\,$K) determining the solar fluorine abundance of $A_\odot(F)=4.40\pm0.25$, which is consistent with the \citet{hall:69} value, with an equally large uncertainty. This nominal value is very close to the meteoritic value, but given the uncertainties in the modelling of the umbral spectra, the value 
to be used is the meteoritic value of \citet{meteoritic:03}. It would be desirable to determine the solar fluorine abundance to a higher accuracy.

\begin{deluxetable*}{lCCCCCCCCC}
\tablecaption{Program stars observed with IGRINS; Stellar parameters and derived abundances in order of \teff\ \label{tab:obs}}
\tablewidth{0pt}
\tablehead{
\colhead{Star}    &  \colhead{\teff} & \colhead{\logg} &
\colhead{\feh} & \colhead{\micro} & \colhead{A(F)} & \colhead{[F/Fe]} & \colhead{A(O)} & \colhead{[O/Fe]} & \colhead{[Ce/Fe]} \\
\nocolhead{ } & \nocolhead{} & \colhead{[K]} &
\colhead{(dex)} & \nocolhead{} & \colhead{[\kms]} &
}
\decimalcolnumbers
\decimals
\startdata
\multicolumn{6}{l}{Stars with a detected HF line:}\\
HIP72012     &  4077 &   1.4 &  -0.28  & 1.5  & 4.30  &   0.15  & 8.58  &  0.17   & -0.06  \\
KIC5113061   &  4100 &   1.7 &  -0.09  & 1.8  & 4.29  &  -0.05  & 8.74  &  0.14   & -0.02 \\
HIP63432     &  4155 &   1.3 &  -0.87  & 1.9  & 3.76  &   0.2   & 8.36  &  0.54   & -0.0 \\
HIP96014     &  4240 &   1.6 &  -0.43  & 1.7  & 3.91  &  -0.09  & 8.48  &  0.22   & -0.05 \\
KIC5779724   &  4303 &   1.6 &  -0.45  & 1.7  & 4.06  &   0.08  & 8.74  &  0.50   & -0.05  \\
KIC4177025   &  4309 &   1.7 &  -0.37  & 1.7  & 4.01  &  -0.05  & 8.75  &  0.43   & -0.09 \\
KIC5113910   &  4338 &   1.7 &  -0.48  & 1.6  & 3.85  &  -0.10   & 8.46  &  0.25  & 0.08     \\
KIC3955590   &  4411 &   2.2 &  +0.03   & 1.6 & 4.67  &   0.21  & 8.89  &  0.17   & -0.07  \\
HD102328     &  4442 &   2.5 &  +0.28   & 1.5 & 4.88  &   0.17  & 8.97  & -0.01   & -0.09    \\
KIC5900096   &  4480 &   2.5 &  +0.23   & 1.5 & 4.78  &   0.12  & 8.94  &  0.02   & -0.09  \\
 & \\
\multicolumn{6}{l}{Stars yielding an upper limit of the HF abundance:}\\
HIP50583          & 4292 &  1.7 &  -0.54 &  1.7 & <3.8  &  < -0.07 & 8.44  & 0.29  &  0.09 \\
KIC11045542       & 4304 &  1.6 &  -0.65 &  1.5 & <3.8  &  <  0.03 & 8.28  & 0.24  & -0.01  \\
$\alpha$ Boo      & 4308 &  1.7 &  -0.55 &  1.8 & <3.9  &  <  0.03 & 8.64  & 0.50  & -0.13    \\
2M14231899        & 4308 &  1.8 &  -0.82 &  1.6 & <3.5  &  < -0.07 & 8.43  & 0.56  &  0.32  \\
2M17215666        & 4342 &  1.6 &  -1.11 &  1.7 & <3.5  &  <  0.13 & 8.14  & 0.56  &  0.34 \\
KIC4659706        & 4428 &  2.5 &  +0.24 &  1.5 & <4.9  &  <  0.23 & 9.03  & 0.10  & -0.04    \\
HIP90344          & 4454 &  2.2 &  -0.39 &  1.4 & <4.3  &  <  0.23 & 8.65  & 0.35  &  0.06 \\
KIC3936921        & 4488 &  2.2 &  +0.01 &  1.6 & <4.5  &  <  0.08 & 8.92  & 0.22  & -0.11 \\
KIC11342694       & 4509 &  2.8 &  +0.14 &  1.3 & <4.7  &  <  0.13 & 8.83  & 0.00  & -0.05   \\
KIC3748585        & 4569 &  2.6 &  +0.03 &  1.3 & <4.6  &  <  0.18 & 8.83  & 0.11  &  0.00  \\
\enddata
\tablecomments{We use A(O)$_\odot = 8.69$ \citep{asplund:09}, A(F)$_\odot = 4.43$ \citep{meteoritic:03}, and A(Ce)$_\odot = 1.58$ \citep{scott:15}.}
\end{deluxetable*}

\begin{deluxetable*}{llCCCCCCCCCCC}
\tablecaption{Program stars observed with the Phoenix spectrograph at KPNO; Stellar parameters and derived abundances\label{tab:phoenix}}
\tablewidth{0pt}
\tablehead{
\colhead{Star} & \colhead{2MASS name} & \colhead{$H$}  & \colhead{$K$}   &  \colhead{\teff} & \colhead{\logg} &
\colhead{\feh} & \colhead{\micro} & \colhead{A(F)} & \colhead{[F/Fe]} & \colhead{A(O)} & \colhead{[O/Fe]} & \colhead{[Ce/Fe]} \\
\nocolhead{ } & \nocolhead{}& \nocolhead{} & \colhead{(mag)}  & \colhead{(mag)} & \colhead{[K]} &
\colhead{(dex)} & \nocolhead{} & \colhead{[\kms]} &
}
\decimalcolnumbers
\decimals
\startdata
HIP48455 & J09524585+2600248 & 1.3   & 1.2  & 4494  & 2.5  & +0.27  & 1.5  & 4.98  &   0.28 & 8.88 & -0.08 & -0.15 \\
HIP68567 & J14021217+4545124 & 3.5   & 3.2  & 4163  & 1.7& -0.17  & 1.5    & 4.20  &  -0.06 & 8.70 &  0.18 & -0.11 \\
HIP69118 & J14085485+3201083 & 5.2   & 5.0  & 4195  & 1.8  & -0.17  & 1.5  & 4.27  &   0.01 & 8.72 &  0.2  & -0.1  \\
HIP69316 & J14111512+3217451 & 3.7   & 3.6  & 4475  & 2.6  & +0.29  & 1.5  & 5.00  &   0.28 & 9.02 &  0.04 & -0.02   \\
HIP70949 & J14304537+0446202 & 2.8   & 2.6  & 4145  & 1.7  & -0.21  & 1.5  & 4.26  &   0.04 & 8.72 &  0.24 & 0.12  \\
HIP72499 & J14492614+1002389 & 4.1   & 4.0  & 4485  & 2.5  & +0.37  & 1.5  & 5.07  &   0.27 & 8.91 & -0.15 & -0.14  \\
HIP73203 & J14574158+2440267 & 3.7   & 3.5  & 4070  & 1.3  & -0.52  & 1.6  & 3.97  &   0.06 & 8.64 &  0.47 & -0.15  \\
HIP73917 & J15062101+2626136 & 4.9   & 4.7  & 4204  & 1.8  & -0.10  & 1.6  & 4.24  &  -0.09 & 8.69 &  0.1  & -0.08  \\
HIP75541 & J15255910+4418079 & 3.9   & 3.9  & 4109  & 1.7  & -0.13  & 1.6  & 4.12  &  -0.18 & 8.71 &  0.15 & -0.02 \\
HIP75572 & J15261738+3420095 & 2.4   & 2.1  & 4014  & 1.3  & -0.43  & 1.6  & 4.04  &   0.04 & 8.52 &  0.26 & -0.09  \\
HIP75583 & J15263014+2807391 & 4.4   & 4.2  & 4169  & 1.6  & -0.41  & 1.5  & 3.98  &  -0.04 & 8.58 &  0.3  & 0.02   \\
HIP76634 & J15390103+0328034 & 4.2   & 4.0  & 4166  & 2.1  & +0.20  & 1.4  & 4.84  &   0.21 & 8.96 &  0.07 & 0.12  \\
HIP77743 & J15522151+2836267 & 5.3   & 5.1  & 4465  & 2.6  & +0.26  & 1.5  & 4.85  &   0.16 & 8.98 &  0.03 & -0.1   \\
HIP78157 & J15573375+1604218 & 5.8   & 5.7  & 4496  & 2.6  & +0.33  & 1.5  & 5.10  &   0.34 & 9.02 &  0.   & -0.02  \\
HIP78262 & J15584908+1612399 & 4.5   & 4.3  & 4070  & 1.7  & -0.05  & 1.5  & 4.39  &   0.01 & 8.80 &  0.16 & -0.14  \\
HIP79120 & J16085888+0327161 & 2.6   & 2.4  & 4106  & 1.8  & +0.10  & 1.4  & 4.69  &   0.16 & 8.90 &  0.11 & -0.11   \\
HIP79488 & J16131544+0501160 & 2.3   & 2.1  & 4067  & 1.6  & -0.11  & 1.7  & 4.24  &  -0.08 & 8.73 &  0.15 & 0.00    \\
HIP79953 & J16191120+4902172 & 2.9   & 2.7  & 4111  & 1.6  & -0.24  & 1.6  & 4.13  &  -0.06 & 8.62 &  0.17 & -0.01 \\
HIP80693 & J16283398+0039540 & 2.2   & 2.0  & 4115  & 1.8  & +0.06  & 1.6  & 4.52  &   0.03 & 8.87 &  0.12 & -0.22   \\
HIP82012 & J16451180+4313015 & 3.0   & 2.6  & 4073  & 1.5  & -0.25  & 1.6  & 4.06  &  -0.12 & 8.60 &  0.16 & 0.02  \\
HIP82611 & J16531756+4724598 & 3.0   & 2.6  & 4163  & 1.6  & -0.47  & 1.6  & 3.80  &  -0.16 & 8.54 &  0.32 & -0.02  \\
HIP82802 & J16552218+1825594 & 2.2   & 2.1  & 4086  & 1.7  & -0.15  & 1.6  & 4.30  &   0.02 & 8.75 &  0.21 & -0.09  \\
HIP83677 & J17060964+0944017 & 3.2   & 2.9  & 4059  & 1.5  & -0.12  & 1.6  & 4.08  &  -0.22 & 8.72 &  0.15 & -0.03  \\
HIP84431 & J17154147+2344338 & 3.2   & 2.8  & 4222  & 1.7  & -0.09  & 1.6  & 4.34  &  -0.   & 8.72 &  0.12 & -0.06   \\
HIP84659 & J17182453+2656130 & 4.8   & 4.6  & 4356  & 2.0  & -0.18  & 1.6  & 4.41  &   0.16 & 8.76 &  0.25 & 0.01  \\
HIP85109 & J17233792+1323514 & 4.5   & 4.3  & 4314  & 2.2  & +0.08  & 1.4  & 4.70  &   0.19 & 8.86 &  0.09 & -0.12   \\
HIP85692 & J17304356+5752365 & 3.0   & 2.8  & 4091  & 1.5  & -0.31  & 1.7  & 3.90  &  -0.22 & 8.55 &  0.17 & 0.18 \\
HIP85824 & J17321358+4619500 & 4.0   & 4.0  & 4170  & 1.7  & -0.25  & 1.5  & 4.31  &   0.13 & 8.81 &  0.37 & -0.03  \\
HIP87445 & J17520472+3958553 & 3.0   & 2.8  & 4158  & 1.6  & -0.26  & 1.6  & 4.06  &  -0.11 & 8.58 &  0.15 & 0.06   \\
HIP87777 & J17555082+2227513 & 3.0   & 2.7  & 4383  & 2.1  & -0.09  & 1.6  & 4.49  &   0.15 & 8.73 &  0.13 & 0.07  \\
HIP88770 & J18072099+0228537 & 3.3   & 3.1  & 4050  & 1.5  & -0.24  & 1.6  & 4.12  &  -0.07 & 8.62 &  0.18 & -0.05  \\
HIP88877 & J18083882+5758468 & 4.1   & 4.0  & 4046  & 1.5  & -0.19  & 1.6  & 4.34  &   0.1  & 8.62 &  0.12 & -0.1    \\
HIP89298 & J18131656+2152493 & 3.0   & 2.8  & 4031  & 1.4  & -0.29  & 1.5  & 4.19  &   0.05 & 8.59 &  0.19 & 0.03   \\
HIP89827 & J18195206+2939588 & 3.3   & 2.9  & 4221  & 1.7  & -0.17  & 1.6  & 4.24  &  -0.02 & 8.62 &  0.1  & 0.00     \\
HIP90915 & J18324614+2337005 & 2.7   & 2.4  & 4008  & 1.5  & -0.18  & 1.8  & 4.24  &  -0.01 & 8.62 &  0.11 & 0.02  \\
HIP92768 & J18541325+2754342 & 2.7   & 2.4  & 4131  & 1.7  & -0.20  & 1.5  & 4.21  &  -0.02 & 8.67 &  0.18 & -0.07 \\
HIP93256 & J18594548+2613492 & 2.2   & 2.0  & 4307  & 1.9  & -0.31  & 1.4  & 4.23  &   0.11 & 8.56 &  0.18 & 0.01   \\
HIP93488 & J19022156+0822248 & 2.7   & 2.5  & 4095  & 1.6  & -0.22  & 1.8  & 4.04  &  -0.17 & 8.66 &  0.19 & 0.14   \\
HIP94591 & J19145845+2823411 & 4.6   & 4.4  & 4181  & 1.6  & -0.21  & 1.6  & 4.01  &  -0.21 & 8.64 &  0.16 & 0.03  \\
HIP96063 & J19315598+3011162 & 4.6   & 4.3  & 4242  & 2.0  & -0.02  & 1.5  & 4.54  &   0.12 & 8.75 &  0.08 & -0.09  \\
HIP97789 & J19521643+3625563 & 3.1   & 2.7  & 4089  & 1.6  & -0.08  & 1.7  & 4.22  &  -0.13 & 8.71 &  0.1  & -0.07 \\
\enddata
\tablecomments{We use A(O)$_\odot = 8.69$ \citep{asplund:09}, A(F)$_\odot = 4.43$ \citep{meteoritic:03}, and A(Ce)$_\odot = 1.58$ \citep{scott:15}.}
\end{deluxetable*}

As can be seen in Table \ref{tab:obs}, only the giants with an effective temperature of less that approximately 4500 K, show the HF line. For the metal-poor stars, even cooler giants are required. 
The method that we use for determining the stellar parameters is developed for stars warmer than approximately 4000 K, which means that we have avoided cooler stars since we aim at high accuracy and homogeneity. We realize that the HF line should get stronger for cooler stars and for stars with lower surface gravities, and these could be used for future fluorine measurements in the metal-poor region \citep[see., e.g., the recent work by][]{rafael:19}. Cool, metal-poor giants are, however, very rare.

\subsection{Uncertainties}

Several effects could contribute to the uncertainties of the determined fluorine abundances. The uncertainties in the HF line data are very small compared to other uncertainties, at least by a factor of ten. Residuals from the telluric line division could impinge on the HF line. We have, however, checked where the telluric lines fall in all of the stars, assuring that this is not the case. Furthermore, the way the continuum is set could affect the equivalent width of the line. We have, however, adjusted the continuum locally, minimizing this uncertainty. The largest uncertainties stem instead from the uncertainties in the stellar parameters of the stars, in spite of our efforts to determine these as accurately as possible. We have therefore allowed the stellar parameters to vary within their uncertainties. The molecular lines are very temperature sensitive, with a change in the derived fluorine of 0.1 dex for $+50$\,K.
A change of the surface gravity of 0.15 dex results in a change of the same magnitude. The uncertainties in abundance ratios are, in general, smaller since they often cancel out to various degrees. A change in the microturbulence does not affect the synthesised line, as expected for these weak lines. This shows that it is important to determine the \teff\ and \logg\ very well in order to minimize the scatter in the fluorine trends with metallicity. We estimate a total uncertainty in the derived fluorine abundances to be $\sigma \mathrm{A(F)} \sim 0.15$\,dex and in the abundance ratios to be $\sigma \mathrm{[F/Fe])} \sim 0.10$\,dex.

The robustness of our method is demonstrated by the similarity between our abundance trends derived from the two different spectrographs (IGRINS and Phoenix), and that determined from the TEXES spectrograph, showing the similarity of the abundances derived from the two diagnostics (the vibration-rotational lines at $2.3$\,\micron\ and the pure rotational lines at $12$\,\micron).


\section{Discussion} \label{sec:disc}

\subsection{Nucleosynthesis of fluorine}\label{nucleo}

In the following a short overview of the possible production sites of fluorine that may be important and are discussed theoretically is given. The predicted trends from the different channels are discussed, which will be important when analyzing our data later on.

{\it (i)} The contributions to the cosmic budget of fluorine from non-rotating massive stars and conventional Type II Supernovae are negligible \citep{Kob:11,prantzos:18}.  
Rapidly rotating massive stars can, however, produce primary fluorine from $^{14}$N, via proton and $\alpha$ captures in the presence of $^{13}$C, which is needed for the generation of protons \citep{prantzos:18}. The $^{14}$N comes from reactions with $^{12}$C which is the ashes of He burning in the massive star itself, and is therefore of primary origin \citep{rafael:19}. This process could dominate the F budget all the way up to solar metallicities. 

{\it ii)} The $\nu$ process, active during core collapse supernovae \citep{woosley:88,woosley:90,timmes:95,langanke:19}, could contribute substantially to the fluorine production \citep{kobayashi:11:nu,nault:13,pila:15}. In this process fluorine is formed from neutrino-induced spallation reactions with $^{20}$Ne in the expelled shell, containing nuclei formed in the progenitor star. Fluorine is then a primary element, formed from a process which is independent of the metallicity of the site of formation. Since the progenitor stars are massive, short-lived stars, they contribute early \citep{olive:19}, already at low metallicities, and at a constant ratio with oxygen, another element synthesized in massive stars ending their lives as supernovae Type II. However, there are still large uncertainties in the stellar modelling (such as progenitor mass and distribution of $^{20}$Ne) and the neutrino-induced thermonuclear reaction rates of $^{20}$Ne \citep{sieverding:18,sieverding:19}. The F production is also very sensitive to the modelled neutrino flux and its spectrum \citep{alibes:01,prantzos:18,olive:19,langanke:19}.

{\it (iii)} During the He-burning thermal pulses (TP) in AGB stars, the $^{14}$N produced in the hydrogen-burning CNO-cycle, can produce fluorine through a chain of reactions also involving neutrons and protons, a process discussed by, for example,  \citet[][]{forestini:92,jorissen:92,abia:11,cristallo:14}. Subsequently, the star undergoes the 3$^\mathrm{rd}$ dredge-up, during which the surface is enriched with fluorine. This fluorine is subsequently expelled to the interstellar medium by stellar winds and/or during the planetary nebulae phase of the star. Fluorine produced through this channel would be a secondary element \citep{prantzos:18}, with yields depending on the metallicity of the star forming it. \citet{goriely:00} also show the importance of partial mixing of protons into the carbon-rich layers in the interiors of AGB stars during the 3$^\mathrm{rd}$ dredge-up, for the formation of fluorine. 


At too high temperatures in the stellar interiors, 
helium-nuclei or proton-capture reactions destroy fluorine, converting it to Ne, as mentioned above. Therefore, the AGB stars that form fluorine are less massive than approximately $4$\,\msun, preventing the high temperatures of hot bottom burning \citep{kobayashi:11:nu}. The fluorine production in AGB stars is, indeed, mass dependent between $2-4$\,\msun, with a maximum around 3\,\msun\ \citep[][]{Kob:11}. Due to the time delay for low- and intermediate-mass stars to start contributing to the cosmic build-up of fluorine, this source will start contributing at a higher metallicity \citep[$\feh\sim-0.9$ to $-0.7$ ][]{kobayashi:11:nu} and will be different for different stellar populations with different star-formation rates. This process was observationally argued for being a dominant source of present-day cosmic fluorine in the solar neighborhood by, for example,  \citet{jorissen:92,recio:12,jonsson:14,olive:19,rafael:19,abia:15:1,abia:15:2}. The last-named also conclude that additional sources are necessary. \citet{abia:19} argue that AGB stars do contribute, however, not as the main source in the solar neighbourhood, thus contrary to \citet{olive:19}, among others. 

{\it (iv)} A similar reaction chain could occur in massive Wolf-Rayet stars, a process investigated early by \citet{meynet:00} and discussed and argued for in, i.a., \citet{cunha:03,renda:04,cunha:08,spitoni:18}, but questioned in \citet{palacios:05} when rotation is included. $^{14}$N, also here resulting from the CNO-cycle as a secondary element (N formed from the pre-existing C), is converted to fluorine during the core-helium burning phase, and subsequently expelled through the strong, metal-line driven wind.  In this channel, the fluorine produced therefore acts as a secondary element, at high enough metallicities. The formation of Wolf-Rayet stars and their winds are also metallicity dependent, thus becoming increasingly important first for higher metallicities \citep{cunha:03} and should, therefore, be less dependent on the star-formation rates of different stellar populations. 
Primary nitrogen, subsequently burning to fluorine, can also be produced, but is restricted to metal-poor, rotating massive stars \citep{meynet:02,spitoni:18}.

It should, however, be noted that the conditions and mechanisms in Wolf-Rayet stars for fluorine to survive and contribute to the cosmic F-reservoir are very uncertain; recent massive-star models with mass-loss rates accounting for the reduction factor due to clumping 
show that these stars might not at all be significant as a source of cosmic fluorine (G. Meynet, private communication). This was also shown earlier by \citet{palacios:05}. 


{\it (v)} \citet{spitoni:18} discuss the possibility of novae contributing to the cosmic budget of fluorine. These could, in principle, produce fluorine \citep{jose:98} through reactions starting with proton captures by $^{17}$O nuclei. Indeed, \citet{spitoni:18} 
conclude that this process might be important to reproduce the secondary behavior of the observed fluorine trends. It should also be noted that the novae yields are also highly uncertain \citep{spitoni:18}.

\begin{figure}[ht!]
\centering
\epsscale{1.00}
\includegraphics[trim={0.2cm 0cm 0cm 0cm},clip,angle=0,width=1.10\hsize]{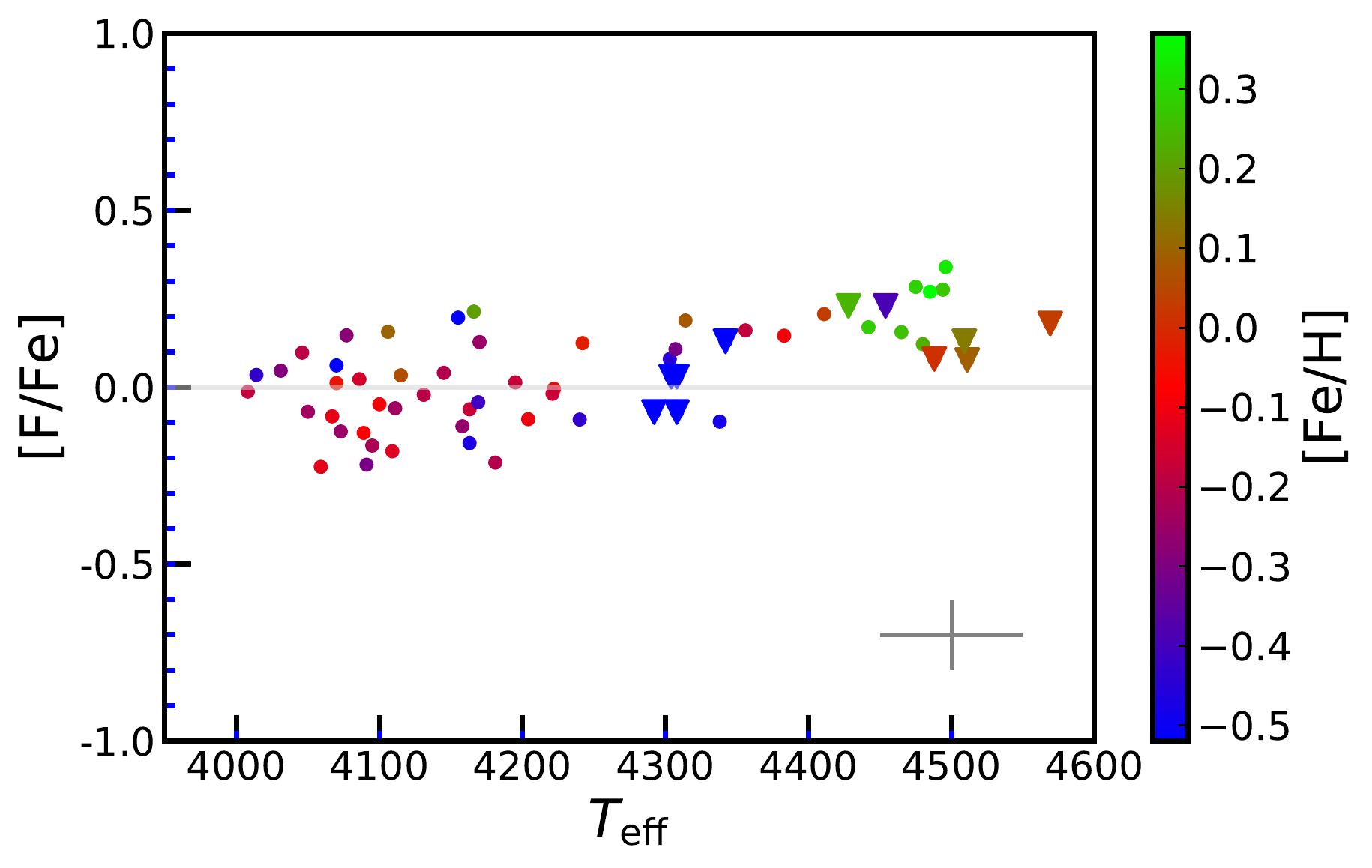}
\caption{[F/Fe] as a function of effective temperature, color-coded for the stars' metallicities, [Fe/H].\label{fig:ffe_vs_teff_feh}}
\end{figure}

\begin{figure*}[ht!]
\centering
\epsscale{1.00}
\includegraphics[trim={0.0cm 0cm 0cm 0cm},clip,angle=0,width=1.00\hsize]{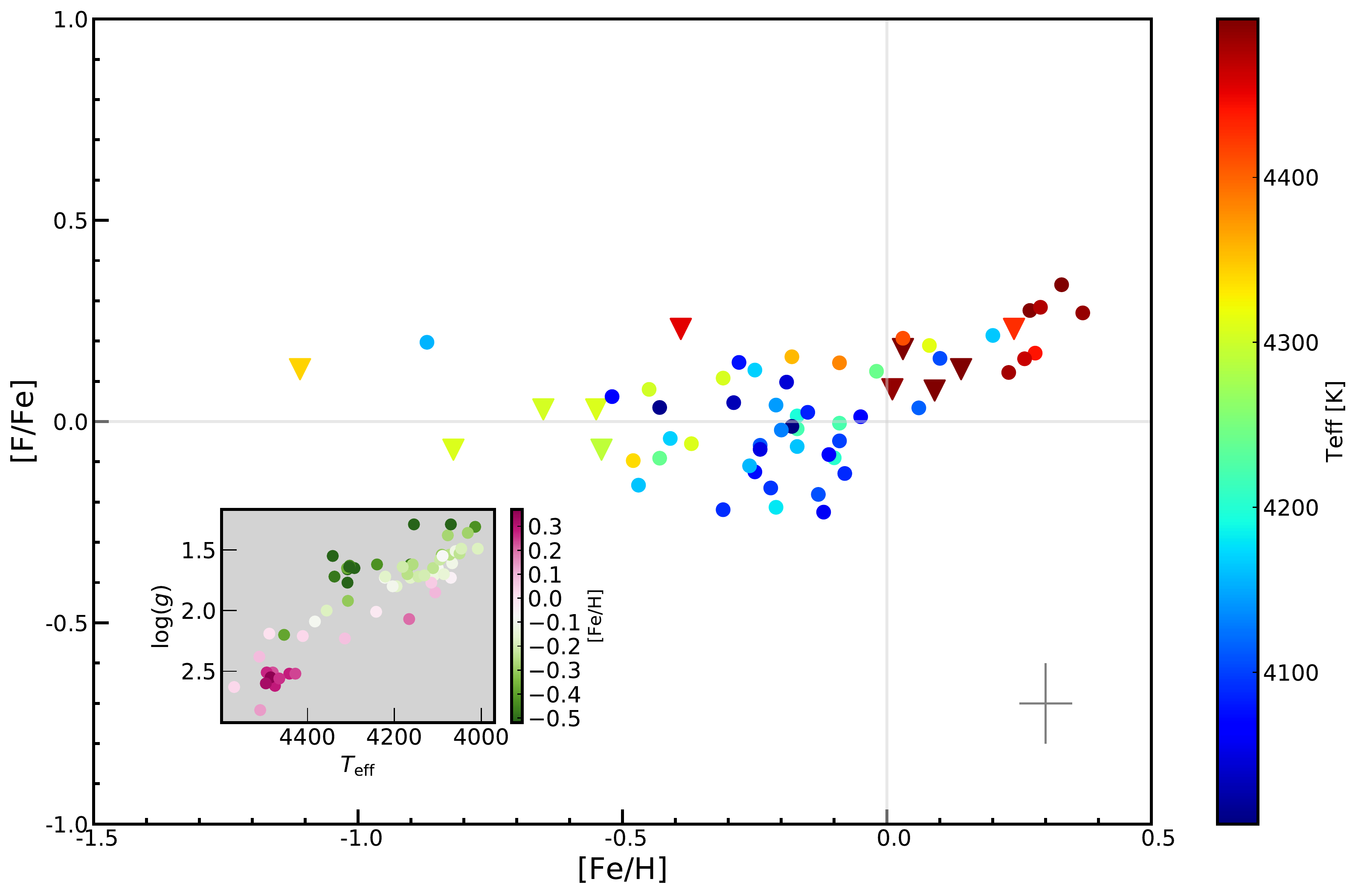}
\caption{[F/Fe] as a function of [Fe/H] color-coded for the stars' effective temperatures, \teff. The figure in the lower left corner is a Kiel-diagram for the same stars color-coded for their metallicities.  \label{fig:ffe_vs_feh_teff}}
\end{figure*}

\subsection{Fluorine abundances and ratios versus metallicity \label{discussion}}

The abundances as a function of metallicity, presented in Figure \ref{fig:fh_vs_feh}, show a very tight relation, close to linear. At lower metallicities these abundance ratios might decrease more slowly as the metallicity decreases. The spread in the F abundances in the range between $\feh\sim-0.6$ up to solar metallicity, is larger than for super-solar metallicities. This is even more evident in Figure \ref{fig:ffe_vs_fe}, where [F/Fe] is plotted versus metallicity. Below $\feh\sim-0.6$ we have three upper limits and one determined abundance from a clearly detected HF line in the spectrum of the giant HIP63432, with $\feh=-0.87$. This spectrum is displayed in Figure \ref{fig:HF} and has a [F/Fe]$=0.2$. Note, that our estimated uncertainties, mainly due to the uncertainties in the stellar parameters, is of the order of $\sigma \mathrm{[F/Fe]} \sim 0.1$\,dex. This low-metallicity trend with metallicity seems, thus, to be quite flat, with indications of a detectable spread. However, more fluorine determinations in metal-poor stars are clearly needed. 

All trends go roughly through the expected solar values, ([F/Fe],$\feh$)=$(0,0)$. It should, however, be kept in mind that the solar F-abundance determinations are subjected to large uncertainties \citep[of the order of $\pm0.25$ dex;][]{hall:69,solar:sme,maiorca:14}, and that we are using the meteoric value of \citet{meteoritic:03}. The choice of the value of the solar fluorine abundance only results in a scaling of the abundances in the Figures, but should be taken into account for comparisons with other published abundances.

Our abundances show less scatter in the trends than many previous investigations \citep[see, e.g,][]{recio:12,pila:15,jonsson:17}.  For the narrow metallicity range of the open cluster giants in \citet{nault:13}, our results agree very well, with a clear increase at super-solar values, see Figure \ref{fig:others}. Furthermore, our new abundances based on the Phoenix spectra show slightly smaller scatter, and a tighter correlation, than those determined by \citet{jonsson:17}, who analysed the same spectra (see the small grey dots in Figure \ref{fig:others}). We believe the reason for this smaller scatter is caused by our more accurate stellar parameters. Also the trends discussed in \citet{rafael:19} agree very well with ours (crosses in Figure \ref{fig:others}), within the common metallicity range. In light of their plateau suggested by the data of \citet{rafael:19} at low metallicities ($\feh<-0.5)$ (excluding the probable members of the Monoceros over-density) and the data point from the study of \citet{li:13}, our indications of a plateau is strengthened. The plateau in \citet{rafael:19}, however, lies at a $\sim0.2$ dex lower level.


In Figure \ref{fig:ffe_vs_teff_feh}, we plot the [F/Fe] abundance ratios as a function of effective temperature, color-coding the stars for their metallicities. Stars cooler than approximately 4350 K show no trend with \teff\ but a scatter. For the warmer stars, only higher [F/Fe] values are measured, tracing the upper range of the scatter.  
This effect is also seen and discussed in \citet{pila:15}. In the warmer stars the HF line increasingly disappears, and only the stars with a high F-abundance will be detected. Lower abundance, such as for metal-poor stars, will not yield a strong enough HF line to be detected for the warm stars. 
Indeed, among the warmer stars, we find mostly metal-rich stars, for which we see an upturn in [F/Fe] in Figure \ref{fig:ffe_vs_fe}. Furthermore, few metal-poor stars are found at temperatures above 4350 K. 

In Figure \ref{fig:ffe_vs_feh_teff}, we  plot the [F/Fe] abundance ratios instead, as a function of  metallicity, colour-coding the stars for their effective temperatures. No trend with \teff\ is detected, apart from the most metal-rich stars being predominately warm. The reason for this is seen in the Kiel diagram in the lower left corner of the figure. There, we clearly see the giant branch with the metallicities increasing diagonally to lower temperatures, as expected \citep[see also][]{jonsson:17I}. Most of our warm stars above 4400 K are red clump stars and are metal-rich. Note that some of the high metallicity stars are also cooler. Had we targetted more stars, we would have detected the HF line in a larger number of cooler stars. The line is actually more easily detected the cooler the giant star is, due to the molecular equilibrium. However, the lower number of warm, metal-poor stars in the Kiel diagram is caused by the HF line strength becoming weaker and eventually disappearing.

\subsection{Thin and thick disk trends}

\begin{figure}[ht!]
\centering
\epsscale{1.00}
\includegraphics[trim={0.2cm 0cm 0cm 0cm},clip,angle=0,width=1.00\hsize]{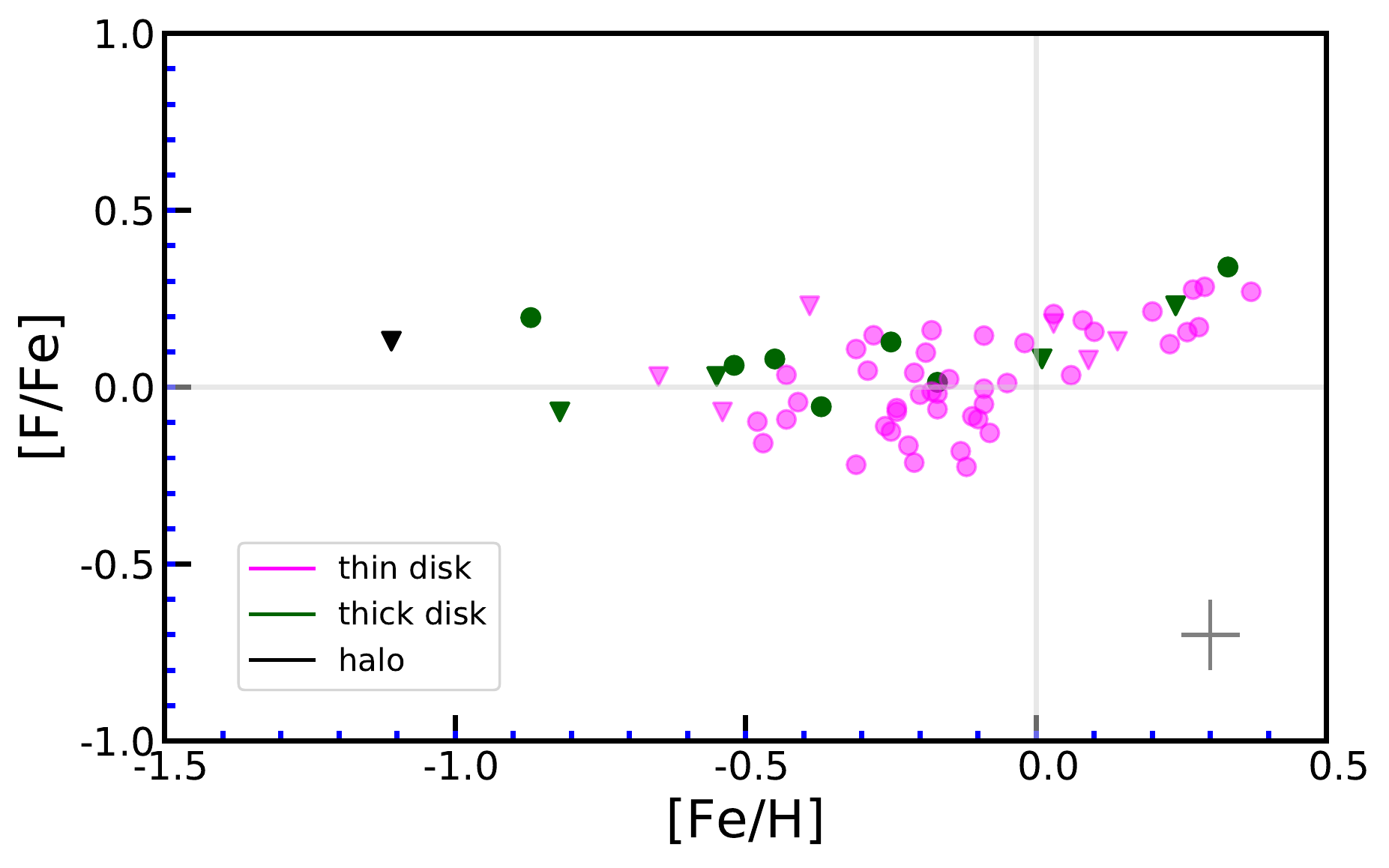}
\caption{[F/Fe] as a function of [Fe/H] for the stars analysed here, color-coded based on their assigned stellar population, i.e. thin disk, thick disk, or halo stars as  determined from the separation in the [Mg/Fe] vs. [Fe/H] plane (see J\"onsson et al. (in prep.) for details)\label{fig:ffe_vs_feh_thin_thick}}.
\end{figure}


In Figure \ref{fig:ffe_vs_feh_thin_thick} we again plot the [F/Fe] ratios as a function of metallicity, but this time colour-coding the stars according to the assigned stellar population, i.e. thin disk, thick disk, or halo stars as  determined from the separation in the [Mg/Fe] vs. [Fe/H] plane (see J\"onsson et al. (in prep.) for details). 
If anything, there is a slight tendency that the thick disk stars seem to lie at the upper envelope of the thin disk stars. More data are needed to confirm this observation. This is, however, theoretically expected according to the models in \citet{Kob:11,kobayashi:11:nu}, also plotted in \citet{jonsson:17}. We have also been able to determine an upper limit of the F abundance for the only halo star in our sample, which is, on the other hand, much lower than what is predicted in \citet{Kob:11}. 

\subsection{Secondary-element behavior}

\begin{figure}[ht!]
\centering
\epsscale{1.00}
\includegraphics[trim={0.2cm 0cm 0cm 0cm},clip,angle=0,width=1.00\hsize]{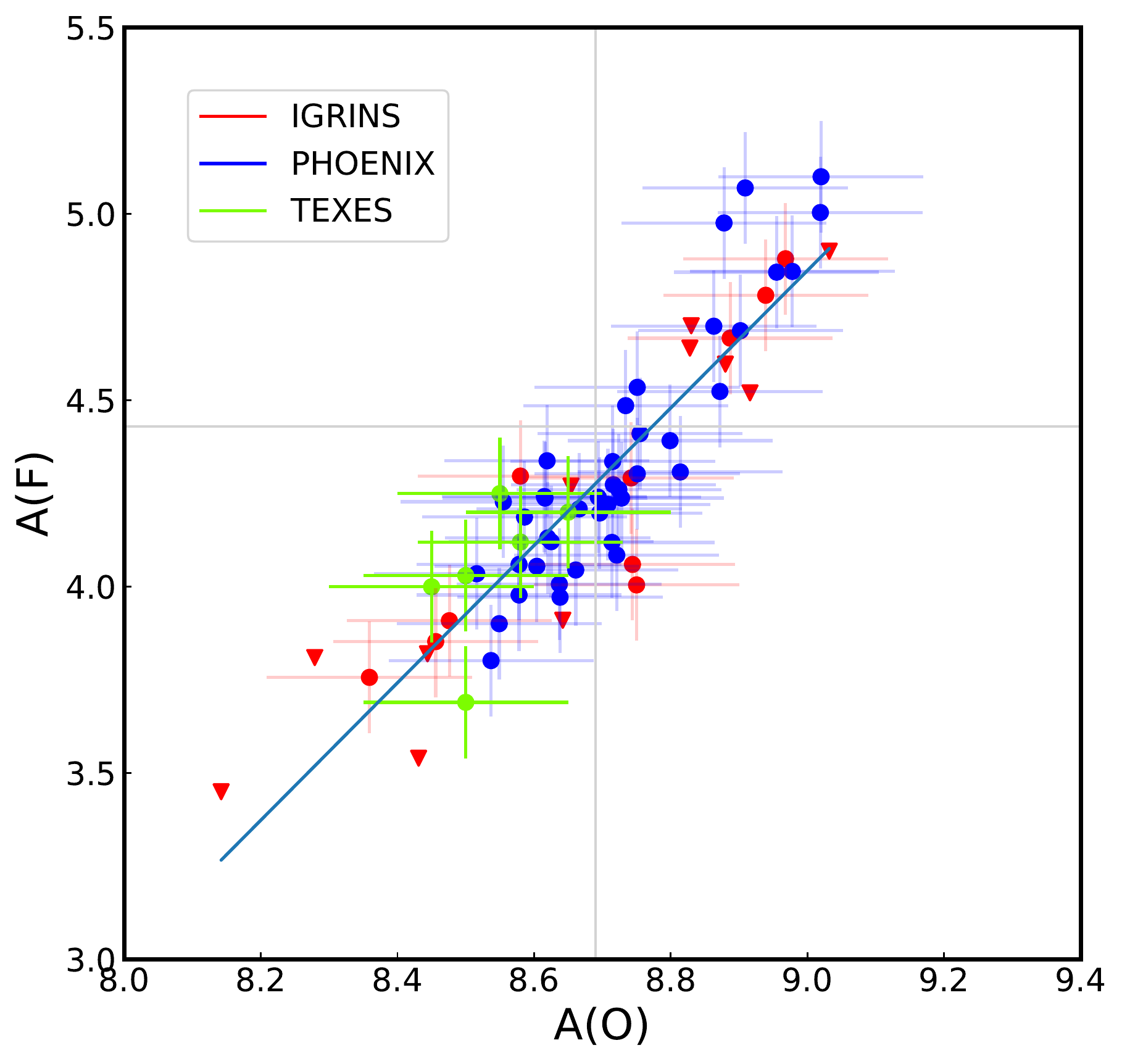}
\caption{Number density of fluorine as a function of the number density of oxygen for the stars analysed here (red and blue) and those from \citet{jonsson:14} from their $12\,\micron$ observations (green). Typical error bars are marked. The blue straight line is a linear regression to the data with a slope of $1.8$, which is close to 2. \label{fig:AF_vs_AO}}
\end{figure}

\begin{figure}[ht!]
\centering
\epsscale{1.00}
\includegraphics[trim={0.2cm 0cm 0cm 0cm},clip,angle=0,width=1.00\hsize]{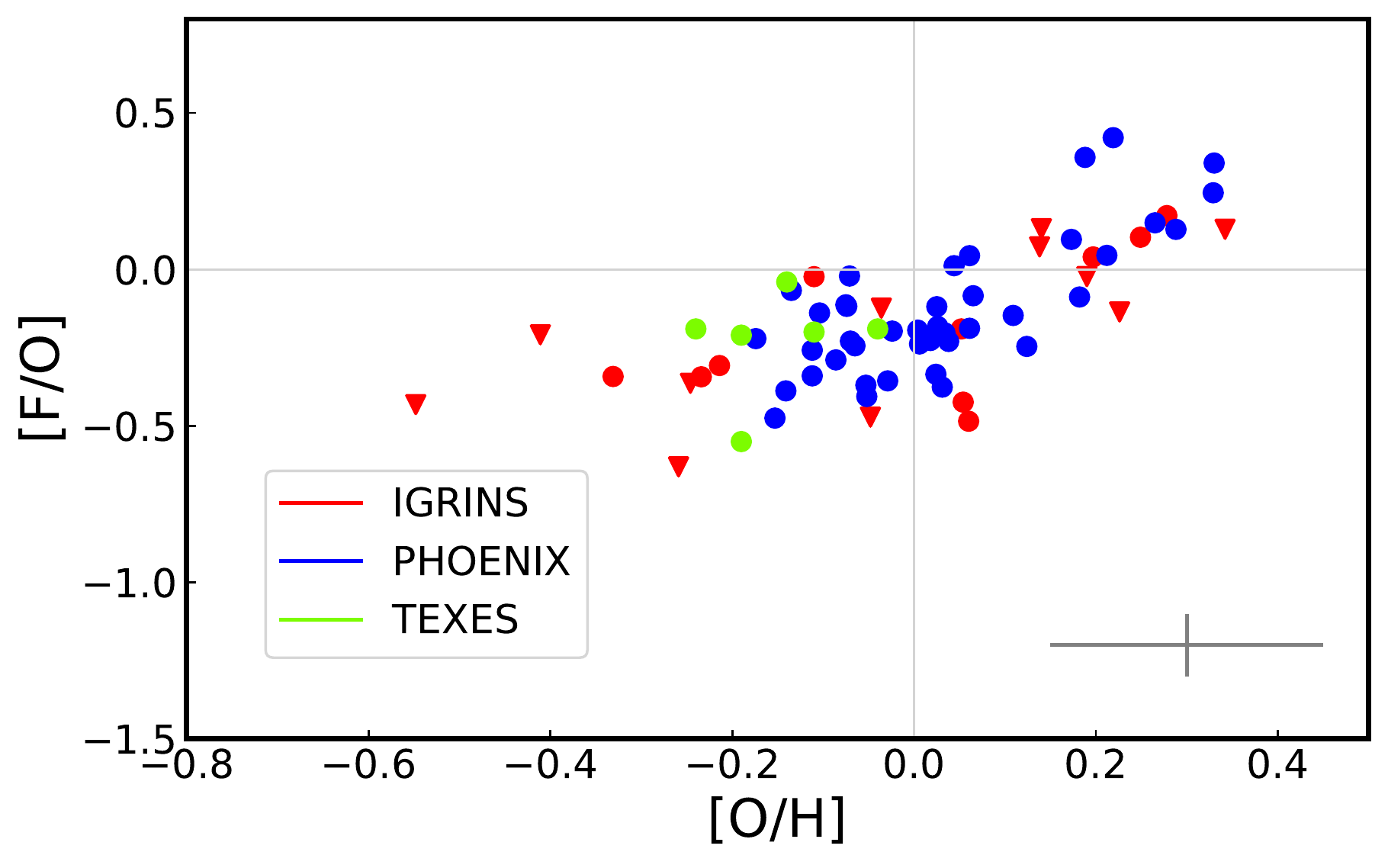}
\caption{Fluorine to oxygen ratio versus the oxygen abundance for the three sets of stars. \label{fig:FO_vs_OH}}
\end{figure}

Another way to investigate the importance of the different nucleosynthetic processes for the cosmic fluorine budget, is to look for the signatures of fluorine's primary or secondary origin, see Section \ref{nucleo}. Since oxygen is produced as a primary element in massive stars \citep[see, e.g.][]{prantzos:18}, plotting the number density of fluorine as a function of the number density of oxygen will reveal whether fluorine is primary or whether it behaves like a secondary element. In Figure \ref{fig:AF_vs_AO} we plot A(F) versus A(O). A linear regression gives a slope of $1.8\pm0.2$ which is a clear indication of a secondary behavior. One can also directly plot the fluorine-to-oxygen abundance ratio as a function of the oxygen abundance, which will be a constant for a primary element and give a slope of one for a secondary element. In Figure \ref{fig:FO_vs_OH} this is plotted and we can reject the  hypothesis of a primary element since the trend is not constant. The slope is instead close to one, $1\pm0.2$, which is the signature of a secondary element.

The synthesis of fluorine in evolved red giant stars has been demonstrated observationally by a number of studies, beginning with that of \citet{jorissen:92} and more recently those of e.g., \citet{pandey:06,alves:11,lucatello:11,abia:15:1,abia:19}. This is the only  nucleosynthetic process forming fluorine that has an observational proof of its activity. From several papers on the galactic chemical evolution of fluorine, the AGB-star contribution in the approximate metallicity range of $-0.6<\feh<0.0$ is, indeed, claimed to be dominant \citep[e.g.][]{renda:04,recio:12,spitoni:18,olive:19}. It should, however,  be noted that there are different theoretical predictions of the AGB-star contributions, mainly due to which nuclear reaction rates are adopted. For instance, \citet{prantzos:18} find a lower AGB stellar yield than \citet{Kob:11} or \citet{spitoni:18}; see also the discussion in \citet{rafael:19}.

Due to the time delay of the AGB stars producing cosmic fluorine, their contribution is theoretically predicted to start contributing at metallicities larger than $\feh\sim-0.9$ for the thin disk and at $\feh\sim-0.7$ for the thick disk \citep[c.f.][]{kobayashi:11:nu,Kob:11,spitoni:18}. Thus, with these theoretical predictions and our observationally determined secondary behaviour, we conclude that the AGB star contribution must be dominant at least in the $-0.6<\feh<0.0$ range, thus corroborating the findings of \citet{recio:12,jonsson:17,olive:19,rafael:19}. The AGB contribution would also naturally cause the large spread in the [F/Fe] versus [Fe/H] plot, which we see in our data, due to the range in masses (and therefore time delays) contributing.

For high metallicities, above solar, we find an increasing [F/Fe]-trend displaying only a small scatter. This part also shows a secondary behaviour.  
The large increasing trend is surprising, since according to galactic chemical evolution models, like the ones in \citet{spitoni:18}, the Fe contribution from Supernovae Type Ia should nominally decrease the trend. The yields needed to increase it instead, are larger than the current best description  of the yields from Wolf-Rayet stars and Novea \citep{spitoni:18}, which are, most likely, even over-estimated. In order to fit the increasing abundances at high metallicities, \citet{spitoni:18} tried to model the galactic chemical evolution of fluorine by artificially increasing the  Wolf-Rayet yields as well as including a fluorine production from novae, which could better model the observed data. Thus, the super-solar fluorine trend can  currently not be explained theoretically.

\subsection{The metal-poor region}

Unfortunately, we do not have many stars in the metal-poor region. We have, however, one robust detection at $\feh=-0.87$ which shows a super-solar [F/Fe] value of $+0.2\pm0.1$. This, together with our upper limits, and in light of the work by \citet{rafael:19} and \citet{li:13}, we cannot reject a (primary) contribution at early times. The AGB-star contribution is much too low at these metallicities. The models predicting a primary contribution at low metallicities are {\it (i)} the rapidly-rotating-massive-star channel \citep{prantzos:18} and {\it (ii)} the $\nu$-process channel \citep[most recently;][]{olive:19}. The trend from the rotating massive stars is a shallow and slightly larger-than-solar [F/Fe] for all metallicities \citep{prantzos:18}, which is in better agreement with our data than the $\nu$-process prediction, which is lower for all metallicities.

\begin{figure}[ht!]
\centering
\epsscale{1.00}
\includegraphics[trim={0.2cm 0cm 0cm 0cm},clip,angle=0,width=1.00\hsize]{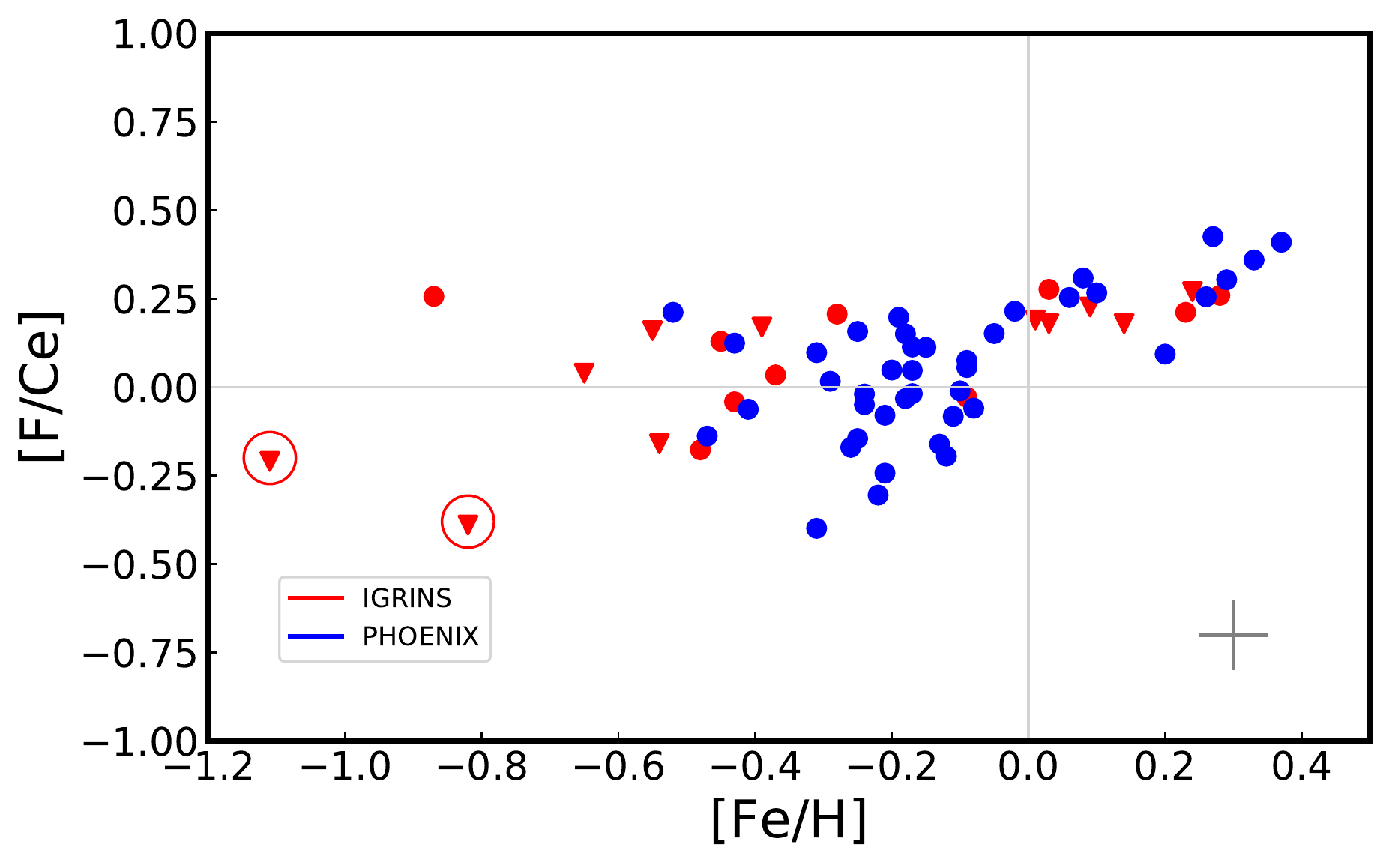}
\caption{Ratio of fluorine to the s-process element Ce which, in general, shows quite a flat trend for sub-solar metallicities. The two giants enhanced in s-process elements are marked with red circles. These have values lower than the expected trend, caused by the high Ce abundances for normal F abundances (upper limits in these cases). The Ce and F abundances are, thus, not enhanced simultaneously in these two stars. We use the solar Ce abundance from \citet{scott:15}: A(Ce)$_\odot = 1.58$.  \label{fig:cef_vs_feh}}
\end{figure}

\subsection{Stars enriched in s-process elements}

The main s-process takes place in the interior of low- and intermediate-mass AGB stars \citep[see, e.g., the discussion in ][]{forsberg:19}, and Ce is to 84\% a main s-process element \citep{bisterzo:14}. The [F/Ce] ratio should then be constant for the interval in metallicities where both elements are predominately synthesized by the same type of AGB stars. We show the [F/Ce] ratio as a function of metallicity in Figure \ref{fig:cef_vs_feh}. Apart from the super-solar metallicities, the ratio is almost flat, although with some scatter. This strengthens the conclusion that AGB stars could be the dominant source of fluorine at least in the $-0.6<\feh<0.0$ range.

In the full sample of stars from which we have chosen our subset of stars, there are a number of stars which are enriched in s-process elements. 
These stars are significantly enriched in Zr, La, and Ce, simultaneously and by the same amount. The origin of these stars and the cause of this enrichment is not known, but being K giants, one possibility could be that the stars are in binary systems, i.e. they are extrinsic s-enhanced stars. The s-process element enhancement is then probably caused by pollution from the evolved AGB companion. This enhancement should then be accompanied by a fluorine enhancement, since fluorine is expected to correlate with the s-process elements, since both production chains include neutrons coming from He-nuclei reactions with $^{13}$C \citep{abia:19}. 
In solar-metallicity carbon stars (on the AGB) this is indeed observed \citep{goriely:00,abia:19}. It would, therefore, be interesting to investigate whether these stars are enhanced in fluorine too.

We have determined upper limits of the fluorine abundances for two of these stars enhanced in s-process elements (2M14231899 and 2M17215666), both below $\feh<-0.8$, one being a thick disk and one a halo star. In Figures \ref{fig:fh_vs_feh} and \ref{fig:ffe_vs_fe}, we see that the upper limits for these stars are not especially high. On the contrary, they are even smaller than the robustly determined fluorine abundance from the star with a metallicity intermediate between the two stars. Thus, the fluorine abundances do not seem abnormally high, although we have very few data points in this metallicity range. The [F/Ce] ratio as a function of metallicity is plotted in Figure \ref{fig:cef_vs_feh}. The high Ce abundance and the normal fluorine abundances put these two stars below the general flat trend, and especially below the ratio of the star with the intermediate metallicity. We thus conclude that these s-enhanced stars do not show abnormally high fluorine abundances. 
This is surprising since it might have been expected, had these elements originated from AGB stars. 

We checked for binarity for these stars, by looking for radial velocity shifts in spectra observed by APOGEE \citep[SDSS IV;][]{apogee:17,apogee:dr14}  at different epochs, but could not find any sign of binarity. Thus, the argument for them being extrinsic s-enhanced stars is weakened. Another possibility may be that they are globular-cluster escapees, which could leave traces of enrichment from AGB stars, and therefore the enhancement of the s-process elements. But then fluorine enhancements could possibly also be expected.

It therefore seems unlikely that AGB stars, at these metallicities, are the cause of the s-process element enhancement or that fluorine is not produced at the same enhancement levels as the s-process elements from AGB stars. At these metallicities, the s-process elements could also have contributions from the r-process, which does not produce fluorine. The fluorine would then be produced in rapidly rotating massive stars or through the $\nu$ process. 

We note that \citet{cunha:03} also found that in their measurements of stars in $\omega$ Cen, stars with enhancements of s-process elements at similar metallicities, revealing a large contribution of AGB stars to the chemical evolution, did not show enhanced levels of fluorine either. These arguments strengthen the non-AGB contribution at lower metallicities. We note, however, also that \citet{abia:19} conclude from analysing carbon-rich AGB stars of different metallicities, that models over-predict the production of fluorine for metal-poor stars ($\feh<-0.5$), which could potentially explain a lower F abundance than expected.
It would be very interesting to measure the fluorine abundances in the other s-process-enriched stars from J\"onsson et al. (in prep.), 
which are at higher metallicities, in a range where the AGB star contribution is high.

We can thus corroborate the results in \citet{rafael:19} who argue for a primary behavior at $\feh<-0.5$ and a  secondary behavior above $-0.5$. 
This is supported by the fact that the s-enhanced stars in our sample do not show enhanced F for metallicities lower than $\feh<-0.6$.

\section{Conclusions}

With the aim of putting constraints on the cosmic origin of fluorine, we have derived the fluorine abundances in 61 K giants in the solar neighborhood with $-1.1<\feh<+0.4$, by analysing the $2.3\,\micron$ HF line. Apart from the Phoenix spectra from \citet{jonsson:17}, we present new spectra observed with the IGRINS spectrograph at high spectral resolution. In order to minimize the scatter in the data, we have carefully determined accurate stellar parameters in a homogeneous way for all these stars. 

We find, in principle, a flat [F/Fe] versus [Fe/H] trend for sub-solar metallicities, but an increasing trend above solar metallicities. 
This increase in [F/Fe] is difficult to explain; even if possible novae and the less likely Wolf Rayet channels are at play for these metallicities, the predicted fluorine production is not enough \citep{spitoni:18}. 

We also find a clear secondary behavior for our stars. In the $-0.6<\feh<0$ range, this secondary behavior, together with the flat [Ce/F] trend, and the theoretically predicted importance of the processes in AGB stars forming cosmic fluorine, leads us to the conclusion that this channel dominates in this metallicity range.

Furthermore, together with the finding that two metal-poor, s-process element-rich stars do not show an enhanced fluorine abundance, and since both fluorine from AGB stars and Fe from SN Ia are time delayed (which for the thick disk means after $\feh\sim-0.7$), 
we cannot reject the hypothesis that for $\feh<-0.7$ the fluorine originates from massive stars, most likely from processes in rapidly rotating massive stars \citep{prantzos:18,rafael:19,olive:19}, showing a primary behavior.

Therefore, it seems likely that several channels are needed to explain the cosmic budget of fluorine at different metallicities, corroborating the discussions in \citet{abia:15:1,cunha:08}. \citet{renda:04} also showed that different contributions are needed. They showed that the $\nu$ process could be dominant at low metallicities, that the significance of the AGB stars successively grows, and that the contribution of Wolf-Rayet stars is significant for solar and super-solar metallicities.  This was also shown by \citet{spitoni:18}. It should, however, be noted that newer models might weaken the case for the Wolf-Rayet channel. \citet{olive:19} also argue that several processes are needed, with the $\nu$ process being important at early times and that the AGB contribution is the major one at solar metallicities. 

It is clear that further observations and measurements of the cosmic fluorine trend at low metallicities are needed. What causes the increasing [F/Fe] ratios for super-solar metallicities also requires further theoretical considerations.

\acknowledgments
We would like to thank George Meynet for fruitful and enlightening discussions on massive-star yields and the theoretically predicted role of Wolf-Rayet stars for the production of fluorine. N.R. acknowledges support from the Swedish Research Council, VR (project numbers 621-2014-5640), and the Royal Physiographic Society in Lund through  the Stiftelse Walter Gyllenbergs fond and M{\"a}rta och Erik Holmbergs donation. 
H. J. acknowledges support from the Crafoord Foundation, Stiftelsen Olle Engkvist Byggm\"astare, and Ruth och Nils-Erik Stenb\"acks stiftelse.
This work used the Immersion Grating Infrared spectrograph (IGRINS) that was developed under a collaboration between the University of Texas at Austin and the Korea Astronomy and Space Science Institute (KASI) with the financial support of the US National Science Foundation under grants AST-1229522 and AST-1702267, of the McDonald Observatory of the University of Texas at Austin, and of the Korean GMT Project of KASI.

%

\facilities{McDonald Observatory (IGRINS), Lowell Observatory (IGRINS), KPNO (Phoenix)}





\vspace{20mm}

\begin{thebibliography}{}
\expandafter\ifx\csname natexlab\endcsname\relax\def\natexlab#1{#1}\fi
\providecommand{\url}[1]{\href{#1}{#1}}
\providecommand{\dodoi}[1]{doi:~\href{http://doi.org/#1}{\nolinkurl{#1}}}
\providecommand{\doeprint}[1]{\href{http://ascl.net/#1}{\nolinkurl{http://ascl.net/#1}}}
\providecommand{\doarXiv}[1]{\href{https://arxiv.org/abs/#1}{\nolinkurl{https://arxiv.org/abs/#1}}}

\bibitem[{{Abia} {et~al.}(2019){Abia}, {Cristallo}, {Cunha}, {de Laverny}, \&
  {Smith}}]{abia:19}
{Abia}, C., {Cristallo}, S., {Cunha}, K., {de Laverny}, P., \& {Smith}, V.~V.
  2019, \aap, 625, A40, \dodoi{10.1051/0004-6361/201935286}

\bibitem[{{Abia} {et~al.}(2015{\natexlab{a}}){Abia}, {Cunha}, {Cristallo}, \&
  {de Laverny}}]{abia:15:1}
{Abia}, C., {Cunha}, K., {Cristallo}, S., \& {de Laverny}, P.
  2015{\natexlab{a}}, \aap, 581, A88, \dodoi{10.1051/0004-6361/201526586}

\bibitem[{{Abia} {et~al.}(2015{\natexlab{b}}){Abia}, {Cunha}, {Cristallo}, \&
  {de Laverny}}]{abia:15:2}
---. 2015{\natexlab{b}}, \aap, 584, C1, \dodoi{10.1051/0004-6361/201526586e}

\bibitem[{{Abia} {et~al.}(2011){Abia}, {Cunha}, {Cristallo}, {de Laverny},
  {Dom{\'\i}nguez}, {Recio-Blanco}, {Smith}, \& {Straniero}}]{abia:11}
{Abia}, C., {Cunha}, K., {Cristallo}, S., {et~al.} 2011, \apjl, 737, L8,
  \dodoi{10.1088/2041-8205/737/1/L8}

\bibitem[{{Alib{\'e}s} {et~al.}(2001){Alib{\'e}s}, {Labay}, \&
  {Canal}}]{alibes:01}
{Alib{\'e}s}, A., {Labay}, J., \& {Canal}, R. 2001, \aap, 370, 1103,
  \dodoi{10.1051/0004-6361:20010296}

\bibitem[{{Alves-Brito} {et~al.}(2011){Alves-Brito}, {Karakas}, {Yong},
  {Mel{\'e}ndez}, \& {V{\'a}squez}}]{alves:11}
{Alves-Brito}, A., {Karakas}, A.~I., {Yong}, D., {Mel{\'e}ndez}, J., \&
  {V{\'a}squez}, S. 2011, \aap, 536, A40, \dodoi{10.1051/0004-6361/201116604}

\bibitem[{{Asplund} {et~al.}(2009){Asplund}, {Grevesse}, {Sauval}, \&
  {Scott}}]{asplund:09}
{Asplund}, M., {Grevesse}, N., {Sauval}, A.~J., \& {Scott}, P. 2009, \araa, 47,
  481, \dodoi{10.1146/annurev.astro.46.060407.145222}

\bibitem[{{Bisterzo} {et~al.}(2014){Bisterzo}, {Travaglio}, {Gallino},
  {Wiescher}, \& {K{\"a}ppeler}}]{bisterzo:14}
{Bisterzo}, S., {Travaglio}, C., {Gallino}, R., {Wiescher}, M., \&
  {K{\"a}ppeler}, F. 2014, \apj, 787, 10, \dodoi{10.1088/0004-637X/787/1/10}

\bibitem[{{Cristallo} {et~al.}(2014){Cristallo}, {Di Leva}, {Imbriani},
  {Piersanti}, {Abia}, {Gialanella}, \& {Straniero}}]{cristallo:14}
{Cristallo}, S., {Di Leva}, A., {Imbriani}, G., {et~al.} 2014, \aap, 570, A46,
  \dodoi{10.1051/0004-6361/201424370}

\bibitem[{{Cunha} {et~al.}(2008){Cunha}, {Smith}, \& {Gibson}}]{cunha:08}
{Cunha}, K., {Smith}, V.~V., \& {Gibson}, B.~K. 2008, \apj, 679, L17,
  \dodoi{10.1086/588816}

\bibitem[{{Cunha} {et~al.}(2003){Cunha}, {Smith}, {Lambert}, \&
  {Hinkle}}]{cunha:03}
{Cunha}, K., {Smith}, V.~V., {Lambert}, D.~L., \& {Hinkle}, K.~H. 2003, AJ,
  126, 1305

\bibitem[{{de Laverny} \& {Recio-Blanco}(2013)}]{delaverny:13}
{de Laverny}, P., \& {Recio-Blanco}, A. 2013, \aap, 560, A74,
  \dodoi{10.1051/0004-6361/201322222}

\bibitem[{{Forestini} {et~al.}(1992){Forestini}, {Goriely}, {Jorissen}, \&
  {Arnould}}]{forestini:92}
{Forestini}, M., {Goriely}, S., {Jorissen}, A., \& {Arnould}, M. 1992, \aap,
  261, 157

\bibitem[{{Forsberg} {et~al.}(2019){Forsberg}, {J{\"o}nsson}, {Ryde}, \&
  {Matteucci}}]{forsberg:19}
{Forsberg}, R., {J{\"o}nsson}, H., {Ryde}, N., \& {Matteucci}, F. 2019, \aap,
  631, A113, \dodoi{10.1051/0004-6361/201936343}

\bibitem[{{Goriely} \& {Mowlavi}(2000)}]{goriely:00}
{Goriely}, S., \& {Mowlavi}, N. 2000, \aap, 362, 599

\bibitem[{{Grevesse} {et~al.}(2007){Grevesse}, {Asplund}, \&
  {Sauval}}]{solar:sme}
{Grevesse}, N., {Asplund}, M., \& {Sauval}, A.~J. 2007, \ssr, 130, 105,
  \dodoi{10.1007/s11214-007-9173-7}

\bibitem[{{Grevesse} {et~al.}(2015){Grevesse}, {Scott}, {Asplund}, \&
  {Sauval}}]{scott:15}
{Grevesse}, N., {Scott}, P., {Asplund}, M., \& {Sauval}, A.~J. 2015, \aap, 573,
  A27, \dodoi{10.1051/0004-6361/201424111}

\bibitem[{{Guer{\c{c}}o} {et~al.}(2019{\natexlab{a}}){Guer{\c{c}}o}, {Cunha},
  {Smith}, {Hayes}, {Abia}, {Lambert}, {J{\"o}nsson}, \& {Ryde}}]{rafael:19}
{Guer{\c{c}}o}, R., {Cunha}, K., {Smith}, V.~V., {et~al.} 2019{\natexlab{a}},
  \apj, 885, 139, \dodoi{10.3847/1538-4357/ab45f1}

\bibitem[{{Guer{\c{c}}o} {et~al.}(2019{\natexlab{b}}){Guer{\c{c}}o}, {Cunha},
  {Smith}, {Pereira}, {Abia}, {Lambert}, {de Laverny}, {Recio-Blanco}, \&
  {J{\"o}nsson}}]{rafael:19:M4}
---. 2019{\natexlab{b}}, \apj, 876, 43, \dodoi{10.3847/1538-4357/ab1340}

\bibitem[{{Gustafsson} {et~al.}(2008){Gustafsson}, {Edvardsson}, {Eriksson},
  {et~al.}}]{marcs:08}
{Gustafsson}, B., {Edvardsson}, B., {Eriksson}, K., {et~al.} 2008, \aap, 486,
  951

\bibitem[{{Hall} \& {Noyes}(1969)}]{hall:69}
{Hall}, D.~N.~B., \& {Noyes}, R.~W. 1969, \aplett, 4, 143

\bibitem[{{Hinkle} {et~al.}(2003){Hinkle}, {Blum}, {Joyce},
  {et~al.}}]{phoenix:2003}
{Hinkle}, K.~H., {Blum}, R.~D., {Joyce}, R.~R., {et~al.} 2003, in Proc. SPIE,
  Vol. 4834, Discoveries and Research Prospects from 6- to 10-Meter-Class
  Telescopes II., ed. P.~{Guhathakurta}, 353

\bibitem[{{Hinkle} {et~al.}(1998){Hinkle}, {Cuberly}, {Gaughan},
  {et~al.}}]{phoenix}
{Hinkle}, K.~H., {Cuberly}, R.~W., {Gaughan}, N.~A., {et~al.} 1998, SPIE, 3354,
  810

\bibitem[{{Holtzman} {et~al.}(2018){Holtzman}, {Hasselquist}, {Shetrone},
  {Cunha}, {Allende Prieto}, {Anguiano}, {Bizyaev}, {Bovy}, {Casey},
  {Edvardsson}, {Johnson}, {J{\"o}nsson}, {Meszaros}, {Smith}, {Sobeck},
  {Zamora}, {Chojnowski}, {Fernandez-Trincado}, {Garcia-Hernandez}, {Majewski},
  {Pinsonneault}, {Souto}, {Stringfellow}, {Tayar}, {Troup}, \&
  {Zasowski}}]{apogee:dr14}
{Holtzman}, J.~A., {Hasselquist}, S., {Shetrone}, M., {et~al.} 2018, {\it AJ},
  156, 125, \dodoi{10.3847/1538-3881/aad4f9}

\bibitem[{{J{\"o}nsson} {et~al.}(2014{\natexlab{a}}){J{\"o}nsson}, {Ryde},
  {Harper}, {Richter}, \& {Hinkle}}]{jonsson:14}
{J{\"o}nsson}, H., {Ryde}, N., {Harper}, G.~M., {Richter}, M.~J., \& {Hinkle},
  K.~H. 2014{\natexlab{a}}, \apjl, 789, L41,
  \dodoi{10.1088/2041-8205/789/2/L41}

\bibitem[{{J{\"o}nsson} {et~al.}(2017{\natexlab{a}}){J{\"o}nsson}, {Ryde},
  {Nordlander}, {Pehlivan Rhodin}, {Hartman}, {J{\"o}nsson}, \&
  {Eriksson}}]{jonsson:17I}
{J{\"o}nsson}, H., {Ryde}, N., {Nordlander}, T., {et~al.} 2017{\natexlab{a}},
  \aap, 598, A100, \dodoi{10.1051/0004-6361/201629128}

\bibitem[{{J{\"o}nsson} {et~al.}(2017{\natexlab{b}}){J{\"o}nsson}, {Ryde},
  {Spitoni}, {Matteucci}, {Cunha}, {Smith}, {Hinkle}, \&
  {Schultheis}}]{jonsson:17}
{J{\"o}nsson}, H., {Ryde}, N., {Spitoni}, E., {et~al.} 2017{\natexlab{b}},
  \apj, 835, 50, \dodoi{10.3847/1538-4357/835/1/50}

\bibitem[{{J{\"o}nsson} {et~al.}(2014{\natexlab{b}}){J{\"o}nsson}, {Ryde},
  {Harper}, {Cunha}, {Schultheis}, {Eriksson}, {Kobayashi}, {Smith}, \&
  {Zoccali}}]{jonsson:14b}
{J{\"o}nsson}, H., {Ryde}, N., {Harper}, G.~M., {et~al.} 2014{\natexlab{b}},
  \aap, 564, A122, \dodoi{10.1051/0004-6361/201423597}

\bibitem[{{Jorissen} {et~al.}(1992){Jorissen}, {Smith}, \&
  {Lambert}}]{jorissen:92}
{Jorissen}, A., {Smith}, V.~V., \& {Lambert}, D.~L. 1992, \aap, 261, 164

\bibitem[{{Jos{\'e}} \& {Hernanz}(1998)}]{jose:98}
{Jos{\'e}}, J., \& {Hernanz}, M. 1998, \apj, 494, 680, \dodoi{10.1086/305244}

\bibitem[{{Kobayashi} {et~al.}(2011{\natexlab{a}}){Kobayashi}, {Izutani},
  {Karakas}, {Yoshida}, {Yong}, \& {Umeda}}]{kobayashi:11:nu}
{Kobayashi}, C., {Izutani}, N., {Karakas}, A.~I., {et~al.} 2011{\natexlab{a}},
  \apjl, 739, L57, \dodoi{10.1088/2041-8205/739/2/L57}

\bibitem[{{Kobayashi} {et~al.}(2011{\natexlab{b}}){Kobayashi}, {Karakas}, \&
  {Umeda}}]{Kob:11}
{Kobayashi}, C., {Karakas}, A.~I., \& {Umeda}, H. 2011{\natexlab{b}}, \mnras,
  414, 3231, \dodoi{10.1111/j.1365-2966.2011.18621.x}

\bibitem[{{Lacy} {et~al.}(2002){Lacy}, {Richter}, {Greathouse}, {Jaffe}, \&
  {Zhu}}]{texes}
{Lacy}, J.~H., {Richter}, M.~J., {Greathouse}, T.~K., {Jaffe}, D.~T., \& {Zhu},
  Q. 2002, {\pasp}, 114, 153

\bibitem[{{Langanke} {et~al.}(2019){Langanke}, {Martinez-Pinedo}, \&
  {Sieverding}}]{langanke:19}
{Langanke}, K., {Martinez-Pinedo}, G., \& {Sieverding}, A. 2019, arXiv
  e-prints, arXiv:1901.03741.
\newblock \doarXiv{1901.03741}

\bibitem[{{Lee} {et~al.}(2017){Lee}, {Gullikson}, \&
  {Kaplan}}]{igrins_pipeline:17}
{Lee}, J.-J., {Gullikson}, K., \& {Kaplan}, K. 2017, {Igrins/Plp 2.2.0},
  Zenodo, \dodoi{10.5281/zenodo.845059}

\bibitem[{{Li} {et~al.}(2013){Li}, {Ludwig}, {Caffau}, {Christlieb}, \&
  {Zhao}}]{li:13}
{Li}, H.~N., {Ludwig}, H.~G., {Caffau}, E., {Christlieb}, N., \& {Zhao}, G.
  2013, \apj, 765, 51, \dodoi{10.1088/0004-637X/765/1/51}

\bibitem[{{Lodders}(2003)}]{meteoritic:03}
{Lodders}, K. 2003, \apj, 591, 1220, \dodoi{10.1086/375492}

\bibitem[{{Lomaeva} {et~al.}(2019){Lomaeva}, {J{\"o}nsson}, {Ryde},
  {Schultheis}, \& {Thorsbro}}]{lomaeva:19}
{Lomaeva}, M., {J{\"o}nsson}, H., {Ryde}, N., {Schultheis}, M., \& {Thorsbro},
  B. 2019, \aap, 625, A141, \dodoi{10.1051/0004-6361/201834247}

\bibitem[{{Lucatello} {et~al.}(2011){Lucatello}, {Masseron}, {Johnson},
  {Pignatari}, \& {Herwig}}]{lucatello:11}
{Lucatello}, S., {Masseron}, T., {Johnson}, J.~A., {Pignatari}, M., \&
  {Herwig}, F. 2011, \apj, 729, 40, \dodoi{10.1088/0004-637X/729/1/40}

\bibitem[{{Mace} {et~al.}(2016){Mace}, {Kim}, {Jaffe}, {Park}, {Lee}, {Kaplan},
  {Yu}, {Yuk}, {Chun}, {Pak}, {Kim}, {Lee}, {Sneden}, {Afsar}, {Pavel}, {Lee},
  {Oh}, {Jeong}, {Park}, {Kidder}, {Lee}, {Nguyen Le}, {McLane},
  {Gully-Santiago}, {Oh}, {Lee}, {Hwang}, \& {Park}}]{mace:16}
{Mace}, G., {Kim}, H., {Jaffe}, D.~T., {et~al.} 2016, Society of Photo-Optical
  Instrumentation Engineers (SPIE) Conference Series, Vol. 9908, {300 nights of
  science with IGRINS at McDonald Observatory}, 99080C,
  \dodoi{10.1117/12.2232780}

\bibitem[{{Mace} {et~al.}(2018){Mace}, {Sokal}, {Lee}, {Oh}, {Park}, {Lee},
  {Good}, {MacQueen}, {Oh}, {Kaplan}, {Kidder}, {Chun}, {Yuk}, {Jeong}, {Pak},
  {Kim}, {Nah}, {Lee}, {Yu}, {Hwang}, {Park}, {Kim}, {Chinn}, {Peck}, {Diaz},
  {Rutten}, {Prato}, {Jacoby}, {Cornelius}, {Hardesty}, {DeGroff}, {Dunham},
  {Levine}, {Nofi}, {Lopez-Valdivia}, {Weinberger}, \& {Jaffe}}]{mace:18}
{Mace}, G., {Sokal}, K., {Lee}, J.-J., {et~al.} 2018, in Society of
  Photo-Optical Instrumentation Engineers (SPIE) Conference Series, Vol. 10702,
  \procspie, 107020Q, \dodoi{10.1117/12.2312345}

\bibitem[{{Maiorca} {et~al.}(2014){Maiorca}, {Uitenbroek}, {Uttenthaler},
  {Randich}, {Busso}, \& {Magrini}}]{maiorca:14}
{Maiorca}, E., {Uitenbroek}, H., {Uttenthaler}, S., {et~al.} 2014, \apj, 788,
  149, \dodoi{10.1088/0004-637X/788/2/149}

\bibitem[{{Majewski} {et~al.}(2017){Majewski}, {Schiavon}, {Frinchaboy},
  {Allende Prieto}, {Barkhouser}, {Bizyaev}, {Blank}, {Brunner}, {Burton},
  {Carrera}, {Chojnowski}, {Cunha}, {Epstein}, {Fitzgerald}, {Garc{\'\i}a
  P{\'e}rez}, {Hearty}, {Henderson}, {Holtzman}, {Johnson}, {Lam}, {Lawler},
  {Maseman}, {M{\'e}sz{\'a}ros}, {Nelson}, {Nguyen}, {Nidever}, {Pinsonneault},
  {Shetrone}, {Smee}, {Smith}, {Stolberg}, {Skrutskie}, {Walker}, {Wilson},
  {Zasowski}, {Anders}, {Basu}, {Beland}, {Blanton}, {Bovy}, {Brownstein},
  {Carlberg}, {Chaplin}, {Chiappini}, {Eisenstein}, {Elsworth}, {Feuillet},
  {Fleming}, {Galbraith-Frew}, {Garc{\'\i}a}, {Garc{\'\i}a-Hern{\'a}ndez},
  {Gillespie}, {Girardi}, {Gunn}, {Hasselquist}, {Hayden}, {Hekker}, {Ivans},
  {Kinemuchi}, {Klaene}, {Mahadevan}, {Mathur}, {Mosser}, {Muna}, {Munn},
  {Nichol}, {O'Connell}, {Parejko}, {Robin}, {Rocha-Pinto}, {Schultheis},
  {Serenelli}, {Shane}, {Silva Aguirre}, {Sobeck}, {Thompson}, {Troup},
  {Weinberg}, \& {Zamora}}]{apogee:17}
{Majewski}, S.~R., {Schiavon}, R.~P., {Frinchaboy}, P.~M., {et~al.} 2017, \aj,
  154, 94, \dodoi{10.3847/1538-3881/aa784d}

\bibitem[{{Meynet} \& {Arnould}(2000)}]{meynet:00}
{Meynet}, G., \& {Arnould}, M. 2000, \aap, 355, 176

\bibitem[{{Meynet} \& {Maeder}(2002)}]{meynet:02}
{Meynet}, G., \& {Maeder}, A. 2002, \aap, 390, 561,
  \dodoi{10.1051/0004-6361:20020755}

\bibitem[{{Nault} \& {Pilachowski}(2013)}]{nault:13}
{Nault}, K.~A., \& {Pilachowski}, C.~A. 2013, \aj, 146, 153,
  \dodoi{10.1088/0004-6256/146/6/153}

\bibitem[{{Olive} \& {Vangioni}(2019)}]{olive:19}
{Olive}, K., \& {Vangioni}, E. 2019, \mnras, 490, 4307,
  \dodoi{10.1093/mnras/stz2893}

\bibitem[{{Palacios} {et~al.}(2005){Palacios}, {Arnould}, \&
  {Meynet}}]{palacios:05}
{Palacios}, A., {Arnould}, M., \& {Meynet}, G. 2005, \aap, 443, 243,
  \dodoi{10.1051/0004-6361:20053323}

\bibitem[{{Pandey}(2006)}]{pandey:06}
{Pandey}, G. 2006, \apjl, 648, L143, \dodoi{10.1086/507888}

\bibitem[{{Pandey} {et~al.}(2008){Pandey}, {Lambert}, \& {Kameswara
  Rao}}]{pandey:08}
{Pandey}, G., {Lambert}, D.~L., \& {Kameswara Rao}, N. 2008, \apj, 674, 1068,
  \dodoi{10.1086/526492}

\bibitem[{{Park} {et~al.}(2014){Park}, {Jaffe}, {Yuk}, {Chun}, {Pak}, {Kim},
  {Pavel}, {Lee}, {Oh}, {Jeong}, {Sim}, {Lee}, {Nguyen Le}, {Strubhar},
  {Gully-Santiago}, {Oh}, {Cha}, {Moon}, {Park}, {Brooks}, {Ko}, {Han}, {Nah},
  {Hill}, {Lee}, {Barnes}, {Yu}, {Kaplan}, {Mace}, {Kim}, {Lee}, {Hwang}, \&
  {Park}}]{igrins:14}
{Park}, C., {Jaffe}, D.~T., {Yuk}, I.-S., {et~al.} 2014, in SPIE, Vol. 9147,
  Ground-based and Airborne Instrumentation for Astronomy V, 91471D,
  \dodoi{10.1117/12.2056431}

\bibitem[{{Pilachowski} \& {Pace}(2015)}]{pila:15}
{Pilachowski}, C.~A., \& {Pace}, C. 2015, \aj, 150, 66,
  \dodoi{10.1088/0004-6256/150/3/66}

\bibitem[{{Prantzos} {et~al.}(2018){Prantzos}, {Abia}, {Limongi}, {Chieffi}, \&
  {Cristallo}}]{prantzos:18}
{Prantzos}, N., {Abia}, C., {Limongi}, M., {Chieffi}, A., \& {Cristallo}, S.
  2018, \mnras, 476, 3432, \dodoi{10.1093/mnras/sty316}

\bibitem[{{Recio-Blanco} {et~al.}(2012){Recio-Blanco}, {de Laverny}, {Worley},
  {Santos}, {Melo}, \& {Israelian}}]{recio:12}
{Recio-Blanco}, A., {de Laverny}, P., {Worley}, C., {et~al.} 2012, \aap, 538,
  A117, \dodoi{10.1051/0004-6361/201118261}

\bibitem[{{Renda} {et~al.}(2004){Renda}, {Fenner}, {Gibson}, {Karakas},
  {Lattanzio}, {Campbell}, {Chieffi}, {Cunha}, \& {Smith}}]{renda:04}
{Renda}, A., {Fenner}, Y., {Gibson}, B.~K., {et~al.} 2004, \mnras, 354, 575,
  \dodoi{10.1111/j.1365-2966.2004.08215.x}

\bibitem[{{Sieverding} {et~al.}(2019){Sieverding}, {Langanke},
  {Mart{\'\i}nez-Pinedo}, {Bollig}, {Janka}, \& {Heger}}]{sieverding:19}
{Sieverding}, A., {Langanke}, K., {Mart{\'\i}nez-Pinedo}, G., {et~al.} 2019,
  \apj, 876, 151, \dodoi{10.3847/1538-4357/ab17e2}

\bibitem[{{Sieverding} {et~al.}(2018){Sieverding}, {Mart{\'\i}nez-Pinedo},
  {Huther}, {Langanke}, \& {Heger}}]{sieverding:18}
{Sieverding}, A., {Mart{\'\i}nez-Pinedo}, G., {Huther}, L., {Langanke}, K., \&
  {Heger}, A. 2018, \apj, 865, 143, \dodoi{10.3847/1538-4357/aadd48}

\bibitem[{{Spitoni} {et~al.}(2018){Spitoni}, {Matteucci}, {J{\"o}nsson},
  {Ryde}, \& {Romano}}]{spitoni:18}
{Spitoni}, E., {Matteucci}, F., {J{\"o}nsson}, H., {Ryde}, N., \& {Romano}, D.
  2018, \aap, 612, A16, \dodoi{10.1051/0004-6361/201732092}

\bibitem[{{Telting} {et~al.}(2014){Telting}, {Avila}, {Buchhave}, {Frandsen},
  {Gandolfi}, {Lindberg}, {Stempels}, {Prins}, \& {NOT staff}}]{fies}
{Telting}, J.~H., {Avila}, G., {Buchhave}, L., {et~al.} 2014, Astronomische
  Nachrichten, 335, 41, \dodoi{10.1002/asna.201312007}

\bibitem[{{Timmes} {et~al.}(1995){Timmes}, {Woosley}, \& {Weaver}}]{timmes:95}
{Timmes}, F.~X., {Woosley}, S.~E., \& {Weaver}, T.~A. 1995, \apjs, 98, 617,
  \dodoi{10.1086/192172}

\bibitem[{{Tody}(1993)}]{IRAF}
{Tody}, D. 1993, in ASP Conf. Ser. 52: Astronomical Data Analysis Software and
  Systems II, ed. R.~J. {Hanisch}, R.~J.~V. {Brissenden}, \& J.~{Barnes}, 173

\bibitem[{{Valenti} \& {Piskunov}(1996)}]{sme}
{Valenti}, J.~A., \& {Piskunov}, N. 1996, \aaps, 118, 595

\bibitem[{{Valenti} \& {Piskunov}(2012)}]{sme_code}
---. 2012, {SME: Spectroscopy Made Easy}.
\newblock \doeprint{1202.013}

\bibitem[{{Werner} {et~al.}(2005){Werner}, {Rauch}, \& {Kruk}}]{werner:05}
{Werner}, K., {Rauch}, T., \& {Kruk}, J.~W. 2005, \aap, 433, 641,
  \dodoi{10.1051/0004-6361:20042258}

\bibitem[{{Woosley} {et~al.}(1990){Woosley}, {Hartmann}, {Hoffman}, \&
  {Haxton}}]{woosley:90}
{Woosley}, S.~E., {Hartmann}, D.~H., {Hoffman}, R.~D., \& {Haxton}, W.~C. 1990,
  \apj, 356, 272, \dodoi{10.1086/168839}

\bibitem[{{Woosley} \& {Haxton}(1988)}]{woosley:88}
{Woosley}, S.~E., \& {Haxton}, W.~C. 1988, \nat, 334, 45,
  \dodoi{10.1038/334045a0}

\bibitem[{{Yuk} {et~al.}(2010){Yuk}, {Jaffe}, {Barnes}, {Chun}, {Park}, {Lee},
  {Lee}, {Wang}, {Park}, {Pak}, {Strubhar}, {Deen}, {Oh}, {Seo}, {Pyo}, {Park},
  {Lacy}, {Goertz}, {Rand}, \& {Gully-Santiago}}]{igrins}
{Yuk}, I.-S., {Jaffe}, D.~T., {Barnes}, S., {et~al.} 2010, in Society of
  Photo-Optical Instrumentation Engineers (SPIE) Conference Series, Vol. 7735,
  Society of Photo-Optical Instrumentation Engineers (SPIE) Conference Series,
  \dodoi{10.1117/12.856864}

\end{thebibliography}
\bibliographystyle{aasjournal}



\end{document}